\documentclass[prl,aps,nofootinbib,notitlepage,twocolumn,10pt]{revtex4-1}

\usepackage{amsmath}
\usepackage{amssymb}
\usepackage{amsfonts}
\usepackage{graphicx}
\usepackage{amsthm}

\usepackage{txfonts}

\usepackage[colorlinks=true,linkcolor=blue,citecolor=blue,urlcolor=blue]{hyperref}

\begin{document}

\title{Sensitivity Bounds for Multiparameter Quantum Metrology}
\author{Manuel Gessner}
\email{manuel.gessner@ino.it}
\affiliation{QSTAR, CNR-INO and LENS, Largo Enrico Fermi 2, I-50125 Firenze, Italy}
\author{Luca Pezz\`e}
\affiliation{QSTAR, CNR-INO and LENS, Largo Enrico Fermi 2, I-50125 Firenze, Italy}
\author{Augusto Smerzi}
\affiliation{QSTAR, CNR-INO and LENS, Largo Enrico Fermi 2, I-50125 Firenze, Italy}
\date{\today}

\begin{abstract}
We identify precision limits for the simultaneous estimation of multiple parameters in multimode interferometers. Quantum strategies to enhance the multiparameter sensitivity are based on entanglement among particles, modes or combining both. The maximum attainable sensitivity of particle-separable states defines the multiparameter shot-noise limit, which can be surpassed without mode entanglement. Further enhancements up to the multiparameter Heisenberg limit are possible by adding mode entanglement. Optimal strategies which saturate the precision bounds are provided.
\end{abstract}

\maketitle

A central problem of quantum metrology is to identify fundamental sensitivity limits and to develop strategies to enhance the precision of parameter estimation~\cite{HelstromBOOK, ParisIJQI2009, Giovannetti2011, TothJPA2014, PezzeRMP2016}. Quantum noise poses an unavoidable limitation even under ideal conditions, in the absence of environmental coupling. Nevertheless, quantum noise can be reduced by adjusting the properties of the probe state and the output measurement. Knowing the sensitivity limits of different classes of probe states is thus crucial to identify quantum resources that lead to an enhancement of sensitivity over classical strategies. The shot noise, i.e., the maximum sensitivity achievable with particle-separable states, and the Heisenberg limit, i.e., the maximum sensitivity achievable with any probe quantum state, have been clearly identified for the estimation of a single parameter~\cite{GiovannettiPRL2006, PezzePRL2009, HyllusPRA2012, PezzePRA2015}. 
Sub-shot-noise sensitivities have been reported in several optical~\cite{Giovannetti2011, KacprowiczNATPHOT2010, KrischekPRL2011} and atomic~\cite{PezzeRMP2016} experiments, opening up strategies to achieve quantum enhancements in matter-wave interferometers~\cite{CroninRMP2009}, 
atomic clocks~\cite{LudlowRMP2015}, quantum sensors~\cite{DegenRMP2017}, gravitational wave detectors \cite{SchnabelNATCOMM2010, AasiNATPHOT2013}, and biological measurements~\cite{Taylor2013}.
However, much less is known about the sensitivity bounds for the simultaneous estimation of multiple parameters. 
What is the shot noise and Heisenberg limit in this case? 
What is the role played by entanglement among the modes where the parameters are encoded? Can multiparticle and multimode entanglement enhance sensitivity?

Multiparameter estimation finds many important applications in quantum imaging \cite{Imaging2007, SpagnoloSCIREP2012,GenoveseJP2016}, 
microscopy and astronomy \cite{Ang2017,Rehacek2017}, sensor networks \cite{KomarNATPHYS2014,Parigi2018},
as well as the detection of inhomogeneous forces, vector fields, and gradients~\cite{Mitchell2011,BaumgratzPRL2016,ApellanizPRA2018}. 
All these tasks go beyond single-parameter estimation.
Only a clear identification of relevant quantum resources can lead to a 
quantum advancement of these technologies~\cite{GenoniPRA2013,HumphreysPRL2013, CiampiniSCIREP2016, GagatsosPRA2016,RagyPRA2016,
ProctorPRL2018, GeARXIV2017, KokPRA2017, ZhuangPRA2018, LiuJPA2016,KokPRA2017, EldredgePRA2018}.

\begin{figure}[tb]
\centering
\includegraphics[width=.49\textwidth]{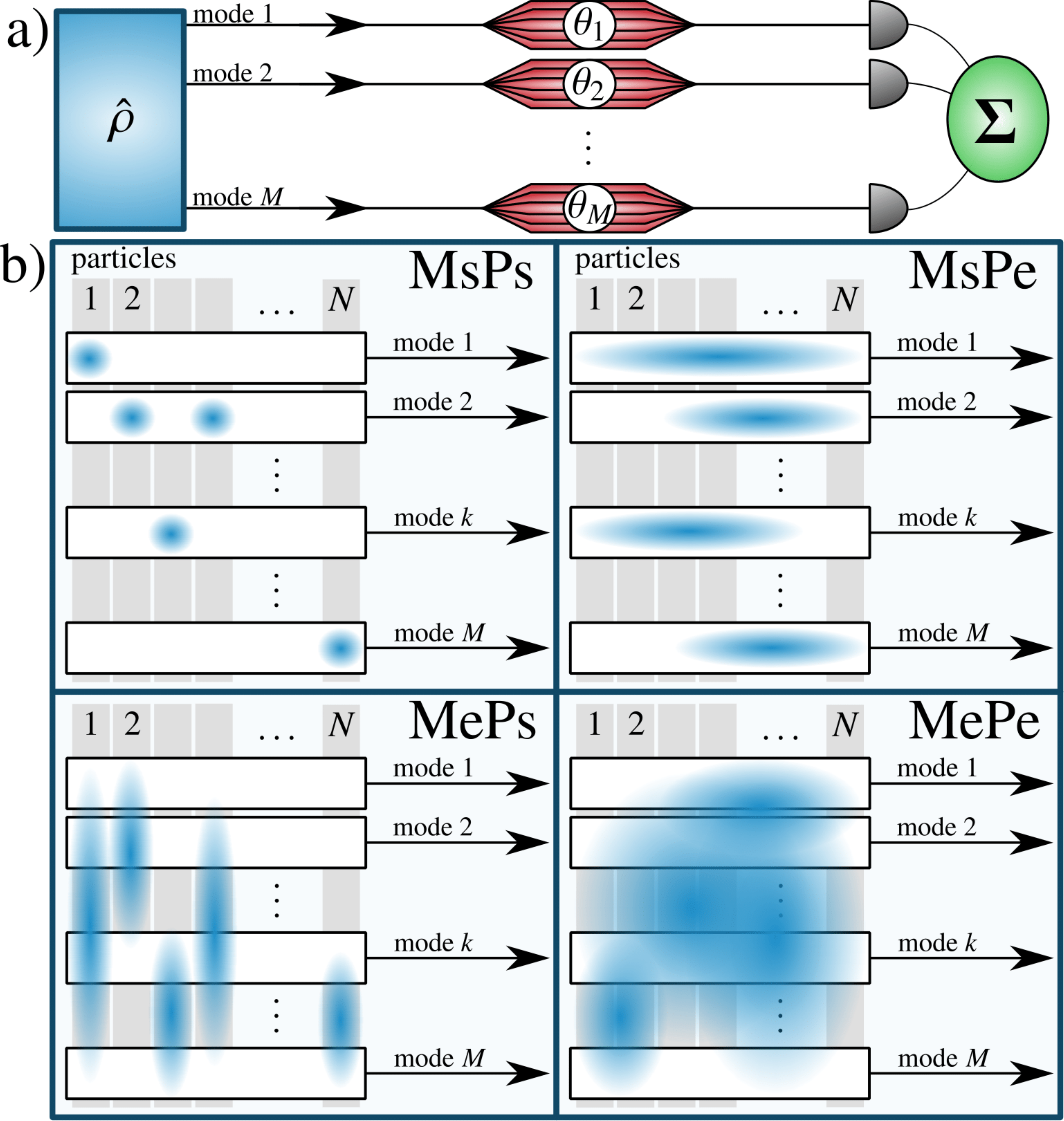}
\caption{General scheme for multiparameter quantum metrology with commuting generators of phase shifts.
(a) The probe state $\hat{\rho}$ of $N$ particles is distributed among $M$ modes. In each mode $k=1,\dots,M$, a parameter $\theta_k$ is encoded as a relative phase shift between sublevels. The sensitivity is quantified by the covariance matrix of the estimators $\mathbf{\Sigma}$. 
The probe state $\hat{\rho}$ can be prepared as schematically shown in (b): mode and particle separable (MsPs), mode separable and particle entangled (MsPe), mode entangled and particle separable (MePs), and mode  and particle entangled (MePe). The grey bars represent the particle partition of the quantum state, the white bars the mode partition. Mode entanglement is illustrated by vertical blue delocalized distributions, particle entanglement by horizontal delocalization.}
\label{fig:1}
\end{figure}

In this manuscript, we present the precision limits for multiparameter quantum metrology in multimode interferometers, see Fig.~\ref{fig:1},
unveiling the nontrivial interplay of mode and particle entanglement. 
The precision limits are given in matrix form, as bounds for the covariance matrix for the estimators of multiple parameters.
As in the single-parameter case, the shot-noise limit is found by maximizing the multiparameter sensitivity over all particle-separable states. 
While particle-separable strategies that use mode entanglement [MePs in Fig.~\ref{fig:1}~b)] can overcome the sensitivity achievable by states that are particle separable and mode separable (MsPs), mode entanglement is not necessary to overcome the multiparameter shot-noise limit. 
The highest sensitivity achievable by mode-separable states is obtained in the presence of particle entanglement (MsPe).
Finally, the multiparameter Heisenberg limit, defined as the sensitivity bound optimized over all quantum states, 
can only be reached if both particle entanglement and mode entanglement (MePe) are present. 
We identify the respective states that saturate the discussed bounds.

\textit{Multimode interferometers for multiphase estimation.---}In the interferometer scheme of Fig.~\ref{fig:1}a),  
each parameter $\theta_k$ is imprinted in one of the $M$ separate mode{s} via the unitary evolution 
$\hat{U}(\boldsymbol{\theta})=\exp(-i\sum_{k=1}^M\hat{H}_k\theta_k)=\exp(- i \hat{\mathbf{H}}\cdot\boldsymbol{\theta})$.
Here, $\boldsymbol{\theta}=(\theta_1,\dots,\theta_M)$ and $\hat{\mathbf{H}}=(\hat{H}_1,\dots,\hat{H}_M)$ are the vectors of unknown 
phases and local Hamiltonians, respectively. 
The initial probe state $\hat{\rho}$ evolves into $\hat{\rho}(\boldsymbol{\theta})=\hat{U}(\boldsymbol{\theta})\hat{\rho} \hat{U}^{\dagger}(\boldsymbol{\theta})$
and it is finally detected. 
We indicate with $\mathbf{x} = (x_1, ..., x_\mu)$ a sequence of $\mu$ independent measurement results that occurs with probability $p(\mathbf{x}|\boldsymbol{\theta}) = \prod_{s=1}^\mu p(x_s |\boldsymbol{\theta})$.
The sensitivity of the multiparameter estimation is determined by the $M\times M$ covariance matrix $\boldsymbol{\Sigma}$ with elements 
$\boldsymbol{\Sigma}_{kl}=\mathrm{Cov}(\theta_{\mathrm{est},k},\theta_{\mathrm{est},l})$, where $\theta_{\mathrm{est},k}(\mathbf{x})$ is 
a locally unbiased estimator for $\theta_k$, with $\langle\theta_{\mathrm{est},k}\rangle=\theta_k$ and 
$d \langle\theta_{\mathrm{est},k}\rangle / \theta_l = \delta_{kl}$~\cite{HelstromBOOK}. 
Any linear combination of the $M$ parameters, $\mathbf{n}\cdot\boldsymbol{\theta}=\sum_{k=1}^Mn_k\theta_k$, is estimated with variance $\Delta^2 (\sum_{k=1}^M n_k\theta_{\mathrm{est},k})=\sum_{kl=1}^Mn_kn_l\mathrm{Cov}(\theta_{\mathrm{est},k},\theta_{\mathrm{est},l})=\mathbf{n}^T\boldsymbol{\Sigma}\mathbf{n}$.
The matrix $\boldsymbol{\Sigma}$ fulfills the chain of inequalities
\begin{align}\label{eq:multiCRLB}
\boldsymbol{\Sigma}\geq \mathbf{F}^{-1}/ \mu\geq \mathbf{F}_Q^{-1}/\mu,
\end{align}
that identify the Cram\'er-Rao (CRB) and quantum Cram\'er-Rao (QCRB) bounds \cite{HelstromBOOK},
respectively, meaning that $\mathbf{n}^T\boldsymbol{\Sigma}\mathbf{n}\geq \mathbf{n}^T\mathbf{F}^{-1}\mathbf{n}/\mu \geq \mathbf{n}^T\mathbf{F}_Q^{-1}\mathbf{n}/\mu$ 
for arbitrary $\mathbf{n}$.
Here $\mathbf{F}^{-1}$ is the inverse of the classical Fisher matrix with elements $\left(\mathbf{F}\right)_{kl}=\sum_xp(x|\boldsymbol{\theta}) \big(\frac{\partial}{\partial\theta_k}\log p(x|\boldsymbol{\theta}) \big) \big(\frac{\partial}{\partial\theta_l}\log p(x|\boldsymbol{\theta}) \big)$, and 
$\left(\mathbf{F}_Q[\hat{\rho}]\right)_{kl} = \mathrm{Tr}[\hat{\rho}\hat{L}_k \hat{L}_l]$, where $d\hat{\rho}/d \theta_k = (\hat{L}_k \hat{\rho}+\hat{\rho}\hat{L}_k)/2$,
are the elements of the quantum Fisher matrix~\cite{HelstromBOOK, ParisIJQI2009}.
$\mathbf{F}$ and $\mathbf{F}_Q$ are positive semi-definite matrices and the chain of inequalities~(\ref{eq:multiCRLB}) is defined only if $\mathbf{F}$ and $\mathbf{F}_Q$ are invertible.
Since in the multimode setting considered here all local Hamiltonians $\hat{H}_k$ commute with each other, the bound $\mathbf{F}=\mathbf{F}_Q$ can always be saturated by an optimally chosen set of local projectors in each mode~\cite{MatsumotoJPA2002, PezzePRL2017}, for instance by the projectors onto the eigenstates of $\hat{L}_k$~\cite{BraunsteinPRL1994}.

We consider probe states of $N$ particles and collective local operators $\hat{H}_k = \sum_{i=1}^N \hat{h}^{(i)}_k$, where $\hat{h}^{(i)}_k$ is a local Hamiltonian for the $i$th particle in the $k$th mode. The $\hat{h}^{(i)}_k$ have the same spectrum $\lambda_{kj}$ with eigenvectors $|\lambda_{kj}^{(i)}\rangle$ for all $i$, where $j$ labels the eigenvalues. For simplicity, we limit the discussion in the main manuscript to the case of two sublevels per mode ($j=\pm$) with $\lambda_{k\pm}=\pm\frac{1}{2}$. A detailed demonstration of all bounds reported below as well as a direct generalization to multilevel systems is given in the Supplementary Material~\cite{Supp}.

\textit{Sensitivity bounds for particle-separable states.---}Here we derive the sensitivity bound for particle-separable states $\hat{\rho}_{\mathrm{p-sep}}=\sum_{\gamma}p_{\gamma}\hat{\rho}^{(1)}_\gamma \otimes\cdots\otimes\hat{\rho}^{(N)}_\gamma$, where $p_{\gamma}$ is a probability distribution and the $\hat{\rho}^{(i)}_\gamma$ are arbitrary single-particle density matrices of the $i$th particle. The quantum Fisher matrix of any particle-separable probe state is bounded by
\begin{align}
\mathbf{F}_Q[\hat{\rho}_{\mathrm{p-sep}},\hat{\mathbf{H}}]\leq 4\sum_{i=1}^N\boldsymbol{\Gamma}[\hat{\rho}^{(i)},\hat{\mathbf{H}}^{(i)}], \notag
\end{align}
where $\boldsymbol{\Gamma}[\hat{\rho}^{(i)},\hat{\mathbf{H}}^{(i)}]$ is the covariance matrix of the reduced density matrix $\hat{\rho}^{(i)}=\sum_{\gamma}p_{\gamma}\hat{\rho}^{(i)}_\gamma$ of particle $i$ with elements $(\boldsymbol{\Gamma}[\hat{\rho}^{(i)},\hat{\mathbf{H}}^{(i)}])_{kl}=\langle \hat{h}^{(i)}_k\hat{h}^{(i)}_l\rangle_{\hat{\rho}^{(i)}}-\langle\hat{h}^{(i)}_k\rangle_{\hat{\rho}^{(i)}}\langle\hat{h}^{(i)}_l\rangle_{\hat{\rho}^{(i)}}$ and $\hat{\mathbf{H}}^{(i)}=(\hat{h}_1^{(i)},\dots,\hat{h}_M^{(i)})$. 
To find the multiparameter shot noise, we maximize $\mathbf{F}_Q[\hat{\rho}_{\mathrm{p-sep}},\hat{\mathbf{H}}]$ over all $\hat{\rho}_{\mathrm{p-sep}}$ with given average particle numbers $\langle\hat{N}_k\rangle$ and $\sum_{k=1}^M \langle\hat{N}_k\rangle = N$. 
We obtain
\begin{align} \label{eq:SN}
\mathbf{F}_{\mathrm{SN}}\equiv\max_{\hat{\rho}_{\mathrm{p-sep}}}\mathbf{F}_Q[\hat{\rho}_{\mathrm{p-sep}},\hat{\mathbf{H}}] = 
\begin{pmatrix} \langle\hat{N}_1\rangle& 0 & \cdots & 0\\
\vdots & &\ddots & \vdots\\
0 & \cdots & 0 & \langle\hat{N}_M\rangle\end{pmatrix}.
\end{align}
The convexity of the quantum Fisher matrix ensures that the bound (\ref{eq:SN}) is achieved by a product of pure single-particle states $|\Psi^{(1)}\rangle\otimes\cdots\otimes|\Psi^{(N)}\rangle$. Optimal states must have the property $\langle\hat{h}_k^{(i)}\rangle_{|\Psi^{(i)}\rangle}=0$ 
for all $k$ and $i$, due to $\lambda_{k+}+\lambda_{k-}=0$, which leads to the diagonal form of $\mathbf{F}_{\mathrm{SN}}$.
If all $\langle \hat{N}_k \rangle>0$, $\mathbf{F}_{\mathrm{SN}}$ is invertible and, according to Eq.~(\ref{eq:multiCRLB}), defines the multiparameter shot-noise limit $\boldsymbol{\Sigma} \geq \boldsymbol{\Sigma}_{\rm SN}/\mu \equiv \mathbf{F}_{\mathrm{SN}}^{-1}/\mu
= {\rm diag}(1/\langle \hat{N}_1 \rangle, 1/\langle \hat{N}_2 \rangle, ..., 1/\langle \hat{N}_M \rangle)/\mu$, i.e., 
the smallest covariance matrix $\boldsymbol{\Sigma}$ for particle-separable probe states. In particular, we recover the shot-noise 
$(\Delta\theta_{\mathrm{est}})^2=1/\mu N$~\cite{GiovannettiPRL2006, PezzePRL2009} in the case of a single parameter ($M=1$).
The shot-noise rank $0\leq r_{\rm SN} \leq M$, defined as the number of positive eigenvalues of the matrix $\mathbf{F}_Q[\hat{\rho},\hat{\mathbf{H}}]-\mathbf{F}_{\mathrm{SN}}$, provides the 
number of linearly independent combinations of the $M$ parameters that can be estimated with sub-shot-noise sensitivity. 
A rank $r_{\rm SN}>0$ can only be achieved by particle-entangled states.

Let us now gain a better understanding of the role of mode entanglement in determining the sensitivity of particle-separable states. 
Considering a pure particle-product state formally corresponds to sending the $N$ particles one-by-one (without any classical correlations) through the $M$-mode interferometer. 
Each of the particles can be localized in a single mode [MsPs strategy depicted in Fig.~\ref{fig:1}~b)], 
or delocalized over several modes (mode entanglement, MePs). We find
\begin{align} \label{ineqTh}
  \mathbf{F}_{\mathrm{MsPs}} \leq 
\mathbf{F}_Q\big[|\Psi^{(1)}\rangle\otimes\cdots\otimes|\Psi^{(N)}\rangle,\hat{\mathbf{H}}\big]
\leq  \mathbf{F}_{\mathrm{MePs}}.
\end{align}
Here $\mathbf{F}_{\mathrm{MePs}}$ is the quantum Fisher matrix obtained by delocalizing each of the particles over all modes according to the weights $p^{(i)}_k=\frac{\langle\hat{N}_k\rangle}{N}$, where $p^{(i)}_k=|\langle\Psi^{(i)}|\lambda_{k+}^{(i)}\rangle|^2+|\langle\Psi^{(i)}|\lambda_{k-}^{(i)}\rangle|^2$ is the probability to find particle $i$ in mode $k$. Moreover, $\mathbf{F}_{\mathrm{MsPs}}$ in Eq.~(\ref{ineqTh}) is the quantum Fisher matrix obtained from fully localized single-particle states, i.e., $p_k^{(i)}=\delta_{kk_i}$ such that $\sum_{i=1}^N\delta_{kk_i}=\langle\hat{N}_k\rangle$, which is only defined for integer $\langle\hat{N}_k\rangle$. In the inequalities~(\ref{ineqTh}) we vary only the distribution of particles among modes, while considering an arbitrary, fixed state preparation within the modes. The result (\ref{ineqTh}) states that, for pure particle-product states, mode entanglement generally leads to a higher sensitivity than strategies based on mode separability. 

Both inequalities in~(\ref{ineqTh}) become equalities for states with the property $\langle\hat{h}_k^{(i)}\rangle_{|\Psi^{(i)}\rangle}=0$ for all $k$ and $i$ and in this case no advantage due to mode entanglement can be achieved. Optimal states that reach the sensitivity limit~(\ref{eq:SN}) are prepared in a balanced superposition of largest and smallest eigenstate within the modes, which ensures that $\langle\hat{h}_k^{(i)}\rangle_{|\Psi^{(i)}\rangle}=0$. Hence, if $\langle\hat{N}_k\rangle$ is integer, we obtain the same sensitivity for the optimal MePs states \cite{nota4}
\begin{align}
|\Psi_{\mathrm{MePs}}\rangle=\bigotimes_{i=1}^N\sum_{k=1}^M\sqrt{\frac{\langle\hat{N}_k\rangle}{2N}}\left(|\lambda_{k+}^{(i)}\rangle+|\lambda_{k-}^{(i)}\rangle\right),\notag
\end{align}
where each particle is delocalized over all modes, and optimal MsPs states
\begin{align}
|\Psi_{\mathrm{MsPs}}\rangle=\bigotimes_{i=1}^N\frac{|\lambda_{k_i+}^{(i)}\rangle+|\lambda_{k_i-}^{(i)}\rangle}{\sqrt{2}},\notag
\end{align}
where each particle is localized on a single mode $k_i$ such that $\sum_{i=1}^N\delta_{kk_i}=\langle\hat{N}_k\rangle$.

\textit{Sensitivity bounds for mode-separable states.---}Let us now determine the upper sensitivity limits for general mode-separable states 
$\hat{\rho}_{\mathrm{m-sep}}=\sum_{\gamma}p_{\gamma}\hat{\rho}_{1,\gamma} \otimes\cdots\otimes\hat{\rho}_{M, \gamma}$, where $\hat{\rho}_{k, \gamma}$ 
is an arbitrary density matrix of mode $k$. The state-dependent bound
\begin{align}\label{eq:MS}
\mathbf{F}_Q[\hat{\rho}_{\mathrm{m-sep}},\hat{\mathbf{H}}]\leq 4 \boldsymbol{\Gamma}[\hat{\rho}_1\otimes\cdots\otimes\hat{\rho}_M,\hat{\mathbf{H}}]
\end{align}
holds, where $\boldsymbol{\Gamma}[\hat{\rho}_1\otimes\cdots\otimes\hat{\rho}_M,\hat{\mathbf{H}}]=\mathrm{diag}((\Delta\hat{H}_1)^2_{\hat{\rho}_1},\dots,(\Delta\hat{H}_M)^2_{\hat{\rho}_M})$ is the covariance matrix of the product state of reduced density matrices $\hat{\rho}_k=\sum_{\gamma}p_{\gamma}\hat{\rho}_{\gamma,k}$ for the different modes $k$~\cite{GessnerPRA2016}. A maximization of the quantum Fisher matrix over all mode-separable states with fixed $\langle\hat{N}_k^2\rangle$ yields:
\begin{align}\label{eq:maxmodesepcov}
\mathbf{F}_{\mathrm{MS}}\equiv\max_{\hat{\rho}_{\mathrm{m-sep}}}\mathbf{F}_Q[\hat{\rho}_{\mathrm{m-sep}},\hat{\mathbf{H}}]=\begin{pmatrix}\langle\hat{N}_1^2\rangle& 0 & \cdots & 0\\
\vdots & &\ddots & \vdots\\
0 & \cdots & 0 &\langle\hat{N}_M^2\rangle\end{pmatrix}.
\end{align}
This sensitivity limit is thus determined by the fluctuations of the number of particles in all modes. 
It should be noticed that $\mathbf{F}_{\mathrm{MS}}\geq \mathbf{F}_{\mathrm{SN}}$ since $\langle\hat{N}_k^2\rangle\geq \langle\hat{N}_k\rangle$. Mode entanglement is therefore not necessary to overcome the multiparameter shot noise. 

For a fixed number of particles $N_k$ in each mode Eq.~(\ref{eq:maxmodesepcov}) reduces to $\mathbf{F}_{\mathrm{MS}}=\mathrm{diag}(N_1^2,\dots,N_M^2)$. The bound is saturated by a product of NOON states,
\begin{align}
|\Psi_{\mathrm{MsPe}}\rangle=\bigotimes_{k=1}^M \frac{ |N_k,+\rangle + |N_k,-\rangle }{\sqrt{2}},\notag
\end{align}
with full $N_k$-particle entanglement in each mode $k$. Here $|N_k,\pm\rangle_k$ describes $N_k$ particles in the state with eigenvalue $\lambda_{k\pm}$. 
In the single-parameter case ($M=1$), the notion of entanglement among different parameter-encoding modes does not exist, and strategies with maximal particle entanglement recover the Heisenberg limit, i.e., $(\Delta\theta_{\mathrm{est}})^2=1/\mu N^2$, achieved by NOON states~\cite{GiovannettiPRL2006, PezzePRL2009}.

Furthermore, for fixed $N_k$, the step-wise enhancement of sensitivity from the bound $\mathbf{F}_{\mathrm{SN}}$ 
for particle-separable states to the bound $\mathbf{F}_{\mathrm{MS}}$ involving full particle entanglement can be probed by deriving bounds for quantum states with a maximal number of entangled particles \cite{HyllusPRA2012} in each mode. Specifically, $\mathbf{P}$-producible states $\hat{\rho}_{\mathbf{P}-\mathrm{prod}}$ are those that contain not more than 
$1\leq P_{k} \leq N_k$ entangled particles in mode $k$ with $\mathbf{P}=\{P_{1},\dots,P_{M}\}$. We obtain $\mathbf{F}^{\mathbf{P}}_{\mathrm{MS}}\equiv\max_{\hat{\rho}_{\mathbf{P}-\mathrm{prod}}}\mathbf{F}_Q[\hat{\rho}_{\mathbf{P}-\mathrm{prod}},\hat{\mathbf{H}}]$ with $\mathbf{F}^{\mathbf{P}}_{\mathrm{MS}}= \mathrm{diag}(s_1P_{1}^2+r_1^2,\dots,s_MP_{M}^2+r_M^2)$, where $s_k=\lfloor N_k/P_{k}\rfloor$ and $r_k=N_k-s_kP_{k}$. 
These bounds are saturated by products of $s_k$ NOON states of $P_k$ particles and a single NOON state of $r_k$ particles in each mode. In general, we obtain the hierarchy 
\begin{align}\label{eq:hierarchyPE}
\mathbf{F}_{\mathrm{MS}}\geq\mathbf{F}^{\mathbf{P}}_{\mathrm{MS}}\geq \mathbf{F}^{\mathbf{P'}}_{\mathrm{MS}}\geq\mathbf{F}_{\mathrm{SN}}
\end{align}
if $P_{k}\geq P'_{k}$ for all $k=1,\dots,M$. We recover $\mathbf{F}_{\mathrm{SN}}$ for $\mathbf{P}=\{1,\dots,1\}$, i.e., in the complete absence of particle entanglement and $\mathbf{F}_{\mathrm{MS}}$ for $\mathbf{P}=\{N_1,\dots,N_M\}$, i.e., maximal particle entanglement in each mode. 

\textit{The multiparameter Heisenberg limit.---}In the following, we identify an ultimate, saturable, lower bound on $\mathbf{n}^T\boldsymbol{\Sigma}\mathbf{n}$ for arbitrary $\mathbf{n}$, minimized over all quantum states. We first derive a weak form of the multiparameter CRB and QCRB,
\begin{align}\label{eq:weakQCRB}
\mathbf{n}^T\boldsymbol{\Sigma}\mathbf{n} \geq \frac{1}{\mu\mathbf{n}^T\mathbf{F}\mathbf{n}}\geq\frac{1}{\mu\mathbf{n}^T\mathbf{F}_Q\mathbf{n}},
\end{align}
respectively, where we chose the normalization $|\mathbf{n}|^2=1$. The inequalities~(\ref{eq:weakQCRB}) can be derived without assuming the existence of the inverse of $\mathbf{F}$ and $\mathbf{F}_Q$ \cite{Supp}. 
While Eq.~(\ref{eq:multiCRLB}) is a matrix inequality and provides bounds for all possible $\mathbf{n}^T\boldsymbol{\Sigma}\mathbf{n} = \Delta^2 (\sum_{k} n_k\theta_{\mathrm{est},k})$ at once, Eq.~(\ref{eq:weakQCRB}) expresses a bound for a single, specific but arbitrary linear combination of 
parameters specified by the vector $\mathbf{n}$~\cite{ProctorPRL2018, GeARXIV2017, EldredgePRA2018}. Since $\mathbf{n}^T\mathbf{A}^{-1}\mathbf{n}\geq(\mathbf{n}^T\mathbf{A}\mathbf{n})^{-1}$ holds for all $\mathbf{n}$ and all matrices 
$\mathbf{A}$, whenever $\mathbf{A}^{-1}$ exists, the chain of inequalities~(\ref{eq:weakQCRB}) is weaker than~(\ref{eq:multiCRLB}). This also means that saturation of the weak bound~(\ref{eq:weakQCRB}) implies saturation of~(\ref{eq:multiCRLB}) whenever it exists.

The state-dependent bound $\mathbf{F}_Q[\hat{\rho},\hat{\mathbf{H}}]\leq 4\boldsymbol{\Gamma}[\hat{\rho},\hat{\mathbf{H}}]$ holds for arbitrary quantum states $\hat{\rho}$, where $\boldsymbol{\Gamma}[\hat{\rho},\hat{\mathbf{H}}]$ is the full covariance matrix. Furthermore, an achievable upper limit on the covariances is given as $\mathbf{n}^T\boldsymbol{\Gamma}[\hat{\rho},\hat{\mathbf{H}}]\mathbf{n}\leq \mathbf{n}^T \boldsymbol{\Gamma}^{\mathbf{n}}[\hat{\rho},\hat{\mathbf{H}}] \mathbf{n}$ for arbitrary $\mathbf{n}$, where $\mathbf{\Gamma}^{\mathbf{n}}[\hat{\rho},\hat{\mathbf{H}}]=\mathbf{v}^{\mathbf{n}}_{\hat{\rho}} \mathbf{v}^{\mathbf{n}T}_{\hat{\rho}}$, and $\mathbf{v}^{\mathbf{n}}_{\hat{\rho}}$ is a vector with elements $\epsilon_k(\Delta \hat{H}_k)_{\hat{\rho}}$, for $k=1,\dots,M$ and $\epsilon_k=\mathrm{sgn}(n_k)$. Maximizing over all quantum states with fixed $\langle\hat{N}_k^2\rangle$ yields $\mathbf{n}^T\mathbf{F}_{\mathrm{HL}}^{\mathbf{n}}\mathbf{n}\equiv\max_{\hat{\rho}}\mathbf{n}^T\mathbf{F}_Q[\hat{\rho},\hat{\mathbf{H}}]\mathbf{n}$ with
\begin{align}\label{eq:HL}
\mathbf{F}_{\mathrm{HL}}^{\mathbf{n}}&=\begin{pmatrix} \langle\hat{N}_1^2\rangle& \cdots & \epsilon_1\epsilon_M\sqrt{\langle\hat{N}_1^2\rangle\langle\hat{N}^2_M\rangle}\\
\vdots &\ddots & \vdots\\
\epsilon_1\epsilon_M\sqrt{\langle\hat{N}_1^2\rangle\langle\hat{N}^2_M\rangle} & \cdots  & \langle\hat{N}_M^2\rangle\end{pmatrix}.
\end{align}
Notice that Eq.~(\ref{eq:HL}) can be written as $\mathbf{F}_{\mathrm{HL}}^{\mathbf{n}}= \mathbf{v}^{\mathbf{n}} \mathbf{v}^{\mathbf{n}T}$, where $\mathbf{v}^{\mathbf{n}}=(\epsilon_1 \sqrt{\langle \hat{N}_1^2 \rangle},\dots,\epsilon_M \sqrt{\langle \hat{N}_M^2 \rangle})$.  
$\mathbf{F}_{\mathrm{HL}}^{\mathbf{n}}$ is a singular rank-one matrix which cannot be inverted:
this implies that the multiparameter Cram\'er-Rao bound~(\ref{eq:multiCRLB}) is not defined while its weaker form~(\ref{eq:weakQCRB}) is.

The multiparameter Heisenberg limit is defined on the basis of Eqs.~(\ref{eq:weakQCRB}) and~(\ref{eq:HL}) as $\mathbf{n}^T\boldsymbol{\Sigma} \mathbf{n}\geq\mathbf{n}^T\boldsymbol{\Sigma}_{\rm HL}^{\mathbf{n}} \mathbf{n}  \equiv (\mu \mathbf{n}^T\mathbf{F}_{\mathrm{HL}}^{\mathbf{n}}\mathbf{n})^{-1}$, and is saturated by the states 
\begin{align}\label{eq:multiNOON}
|\Psi^{\mathbf{n}}_{\mathrm{MePe}}\rangle&=\frac{1}{\sqrt{2}}(|N_1,\epsilon_1\rangle\otimes|N_2,\epsilon_2\rangle\otimes\dots\otimes|N_M,\epsilon_M\rangle\\&\qquad+|N_1,-\epsilon_1\rangle\otimes|N_2,-\epsilon_2\rangle\otimes\dots\otimes|N_M,-\epsilon_M\rangle)\notag,
\end{align}
for arbitrary $\mathbf{n}$. Both the states~(\ref{eq:multiNOON}) and the matrix~(\ref{eq:HL}) depend on the sign of the components of $\mathbf{n}$. The states~(\ref{eq:multiNOON}) contain entanglement among all modes and among all of the $N_k$ particles in each mode. In the single-mode case $(M=1)$ this reduces to the standard NOON state and we again recover the Heisenberg limit $(\Delta \theta_{\mathrm{est}})^2 = 1/\mu N^2$.

\textit{Sensitivity bounds for separability among specific modes.---}To probe the transition from complete mode separability to full $M$-mode entanglement, we derive bounds for quantum states that contain entanglement only between specific subsets of the $M$ modes. States that are mode separable in the partition $\Lambda=\mathcal{A}_1|\dots|\mathcal{A}_L$, where the $\mathcal{A}_m$ describe groups of modes, can be written as $\hat{\rho}_{\Lambda-\mathrm{sep}}=\sum_{\gamma}p_{\gamma}\hat{\rho}_{\gamma,\mathcal{A}_1}\otimes\dots\otimes\hat{\rho}_{\gamma,\mathcal{A}_L}$, with density matrices $\hat{\rho}_{\gamma,\mathcal{A}_m}$ on $\mathcal{A}_m$. Following~\cite{Quantum2017,Qin2018}, we obtain the state-dependent upper bound 
\begin{align}
\mathbf{F}_Q[\hat{\rho}_{\Lambda-\mathrm{sep}},\hat{\mathbf{H}}]\leq 4 \boldsymbol{\Gamma}[\hat{\rho}_{\mathcal{A}_1}\otimes\cdots\otimes\hat{\rho}_{\mathcal{A}_L},\hat{\mathbf{H}}],\notag
\end{align}
where $\hat{\rho}_{\mathcal{A}_m}=\sum_{\gamma}p_{\gamma}\hat{\rho}_{\gamma,\mathcal{A}_m}$ is the reduced density matrix for $\mathcal{A}_m$. This matrix is obtained from the full covariance matrix $\boldsymbol{\Gamma}[\hat{\rho}_{\Lambda-\mathrm{sep}},\hat{\mathbf{H}}]$ by removing all off-diagonal elements that describe correlations between the $\mathcal{A}_m$ while retaining the correlations within each of the $\mathcal{A}_m$. 

By combining the methods used for the derivation of Eqs.~(\ref{eq:MS}) and~(\ref{eq:HL}), the sensitivity limits $\mathbf{F}^{\mathbf{n}}_{\Lambda}$ for the states $\hat{\rho}_{\Lambda-\mathrm{sep}}$ can be obtained. The result is obtained from $\mathbf{F}_{\mathrm{HL}}^{\mathbf{n}}$ by setting to zero the off-diagonal elements that describe mode correlations across different groups $\mathcal{A}_m$. These matrices interpolate between the sensitivity limits of fully $M$-mode entangled states $\mathbf{F}_{\mathrm{HL}}^{\mathbf{n}}$ and fully mode separable states $\mathbf{F}_{\mathrm{MS}}$. This is expressed by the hierarchy
\begin{align}\label{eq:hierarchyME}
\mathbf{n}^T\mathbf{F}^{\mathbf{n}}_{\mathrm{HL}}\mathbf{n} \geq \mathbf{n}^T\mathbf{F}^{\mathbf{n}}_{\Lambda_A}\mathbf{n}\geq \mathbf{n}^T\mathbf{F}^{\mathbf{n}}_{\Lambda_B}\mathbf{n}\geq \mathbf{n}^T\mathbf{F}_{\mathrm{MS}}\mathbf{n},
\end{align}
which holds for all $\mathbf{n}$ and any pair of partitions $\Lambda_A,\Lambda_B$, such that the subsets in $\Lambda_A$ can be obtained by joining subsets of $\Lambda_B$. 
The sensitivity $\mathbf{F}^{\mathbf{n}}_{\Lambda}$ can be reached by mode products of states of the form~(\ref{eq:multiNOON}) for each of the $\mathcal{A}_m$. For a fixed number of particles, the lowest (fully mode separable) bound in~(\ref{eq:hierarchyME}) constitutes the largest bound in the hierarchy~(\ref{eq:hierarchyPE}) as a function of the number of entangled particles.

\textit{Enhancement of sensitivity by multimode and multiparticle entanglement.---}The role of mode entanglement for quantum multiparameter estimation 
has been studied intensively over recent years \cite{GenoniPRA2013,HumphreysPRL2013, CiampiniSCIREP2016, GagatsosPRA2016, RagyPRA2016,
ProctorPRL2018, GeARXIV2017, KokPRA2017, ZhuangPRA2018, LiuJPA2016,KokPRA2017, EldredgePRA2018}. 
No general consensus on the possible advantage of mode entanglement has been reached. 
Many studies have focused their analysis to the sum $\sum_{k=1}^M(\Delta\theta_{\mathrm{est},k})^2$ of single-parameter sensitivities or the weighted sum $\sum_{k=1}^M w_k^2 (\Delta\theta_{\mathrm{est},k})^2$ with $w_k\geq 0$. Both these figures of merit ignore possible correlations between the parameters and lead to the result that mode correlations can only have a detrimental influence on the sensitivity. This can be seen by taking the trace on the QCRB~(\ref{eq:multiCRLB}), $\sum_{k=1}^M(\Delta\theta_{\mathrm{est},k})^2\geq\sum_{k=1}^M(\mathbf{F}_Q^{-1})_{kk}$, which is always larger or equal to the sum of single-parameter sensitivities $\sum_{k=1}^M(\mathbf{F}_Q)^{-1}_{kk}$ (see, e.g., \cite{Kay1993}). Mode entanglement establishes correlations that can lead to an enhancement of phase sensitivity only when considering a figure of merit that includes the covariances among the parameters. 
This possibility is fully accounted for when studying bounds for $\boldsymbol{\Sigma}$ in full matrix form, as done in this manuscript.

The figure of merit $\mathbf{n}^T\boldsymbol{\Sigma}\mathbf{n}=\sum_{kl=1}^Mn_kn_l\mathrm{Cov}(\theta_{\mathrm{est},k},\theta_{\mathrm{est},l})$ may include covariances between the parameters, in addition to the weighted sum of single-parameter variances. 
Let us illustrate the quantum gain due to multimode and multiparticle entanglement in~(\ref{eq:weakQCRB}) using the example of an equally weighted linear combination of parameters, $|n_k|=1/\sqrt{M}$ with arbitrary signs, and an equal and integer number of $N_k =\bar{N} =N/M$ particles in each mode. We determine the maximal sensitivity $S^{\max}_{M_e,P_e}=\max_{\hat{\rho}_{M_e,P_e}}\mathbf{n}^T\mathbf{F}_Q[\hat{\rho}_{M_e,P_e},\hat{\mathbf{H}}]\mathbf{n}$ for quantum states $\hat{\rho}_{M_e,P_e}$ with up to $P_e\leq N/M$ entangled particles in each mode and up to $M_e\leq M$ entangled modes. Notice that $P_e=1$ does not necessarily imply full particle separability since it only demands that there is no entanglement among the particles that enter the same mode. If additionally $M_e=1$, we have a fully mode- and particle-separable state with shot-noise sensitivity $S^{\max}_{1,1}=N$. The gain factor $G_{M_e,P_e}=S^{\max}_{M_e,P_e}/S^{\max}_{1,1}=(sP_e^2+r^2)(uM_e^2+v^2)/(NM)$ expresses the largest achievable quantum-enhancement over the shot-noise limit, where $s=\lfloor \bar{N}/P_e\rfloor$, $r=\bar{N}-sP_e$, $u=\lfloor M/M_e\rfloor$ and $v=M-uM_e$. Special cases of interest are given by
\begin{eqnarray}
G_{1,1}=\:&1, \quad G_{1,\bar{N}}&\:=\bar{N},\notag\\
G_{M,1}=\:&M, \quad G_{M,\bar{N}}&\:=\bar{N}M.\notag
\end{eqnarray}
We observe that local particle entanglement in each mode can achieve an enhancement of up to $\bar{N}$ (corresponding to the number of entangled particles per mode) while mode entanglement can increase the sensitivity by a factor of $M$ (corresponding to the number of entangled modes). By combining both, we can achieve a gain factor up to $\bar{N}M$.

Finally, we remark that our results can be extended to provide bounds on more general figures of merit $\mathrm{Tr}\{\mathbf{W}\boldsymbol{\Sigma}\}$, where $\mathbf{W}\geq 0$ is an arbitrary weight matrix. The sensitivity bounds and optimal states are obtained by performing a mode transformation that diagonalizes the matrix $\mathbf{W}$ \cite{Supp}.

\textit{Conclusions.---}We identified sensitivity bounds and optimal states for the simultaneous estimation of multiple parameters in multimode interferometers and characterized the interplay between mode and particle entanglement. Our bounds are given in terms of the full Fisher matrix and are valid for any linear combination of estimators taking into account correlations between parameters. In particular, this led to the identification of the multiparameter shot-noise limit in matrix form -- corresponding to the maximum sensitivity achievable by particle-separable states -- and the Heisenberg limit -- corresponding to the maximum sensitivity achievable for any probe state. Particle entanglement is thus necessary to overcome the multiparameter shot-noise limit with a fixed number of probe particles. When correlations between the parameters are present, the multiparameter sensitivity further grows with the number of entangled modes. This reveals the possibility to achieve a collective quantum-enhancement for the estimation of multiple parameters beyond an optimized point-by-point estimation of individual parameters. 

Our results build the foundation for the development of genuine quantum technological strategies in applications that rely on the precise acquisition of an ensemble of parameters, such as sensing of spatially distributed fields and imaging techniques. Experimental realizations are possible with existing technology in a wide range of atomic and photonic systems that provide coherent access to multiple modes, see, e.g., \cite{PezzeRMP2016,SpagnoloSCIREP2012,Parigi2018,Mitchell2011,CiampiniSCIREP2016,Qin2018}.

\textit{Acknowledgments.---}M.G. acknowledges funding by the Alexander von Humboldt foundation.
This work has been supported by the European Commission through the QuantERA projects ``Q-Clocks" and ``CEBBEC".

\clearpage

\begin{center}
{\large \bfseries Supplementary Material}
\end{center}

\setcounter{equation}{0} \setcounter{figure}{0} \renewcommand\thefigure{S%
\arabic{figure}} \renewcommand\theequation{S\arabic{equation}}



\section{General framework} 
\label{General}

\subsection{Mode and particle representations of the interferometer}
\label{Interf}
We study multiparameter interferometers, where each parameter is imprinted in a separate set of modes and there are no interactions among the particles in the interferometer. The evolution is therefore local both in the modes and particles. In the following we discuss the two corresponding representations.

\subsubsection{Mode representation}
\label{sec:mrep}
The phases $\boldsymbol{\theta}=(\theta_1,\dots,\theta_M)$ are imprinted in separate modes through the unitary evolution
\begin{align} \label{eq:modeevolution}
U(\boldsymbol{\theta})=\exp\left(-i\hat{\mathbf{H}}\cdot\boldsymbol{\theta}\right)=\exp\left(-i\sum_{k=1}^M\hat{H}_k\theta_k\right), 
\end{align}
where $\hat{\mathbf{H}}=(\hat{H}_1,\dots,\hat{H}_M)$ and $\hat{H}_k$ are local Hamiltonians for the modes $k=1,\dots,M$. 
Each of the Hamiltonians $\hat{H}_k$ acts on a separate mode Hilbert space $ \mathcal{H}_k$ and 
we may describe the full Hilbert space by the tensor product $\mathcal{H}= \mathcal{H}_1 \otimes\cdots\otimes\mathcal{H}_M$ (Fig.~\ref{fig:modevspartA} a). The spectral decomposition of the mode Hamiltonian $\hat{H}_k$ is $\hat{H}_k = \sum_j \Lambda_{kj} \vert \Lambda_{kj} \rangle \langle \Lambda_{kj} \vert$, where $\Lambda_{kj}$ and $\vert \Lambda_{kj} \rangle$ are eigenvalues and corresponding eigenvectors of $\hat{H}_k$, respectively,
and the completeness relation is $\sum_j \vert \Lambda_{kj} \rangle \langle \Lambda_{kj} \vert = \mathbb{I}_k$, where $\mathbb{I}_k$ is the identity on $\mathcal{H}_k$.

\subsubsection{Particle representation}
\label{sec:prep}
\begin{figure}[tb]
\centering
\includegraphics[width=.49\textwidth]{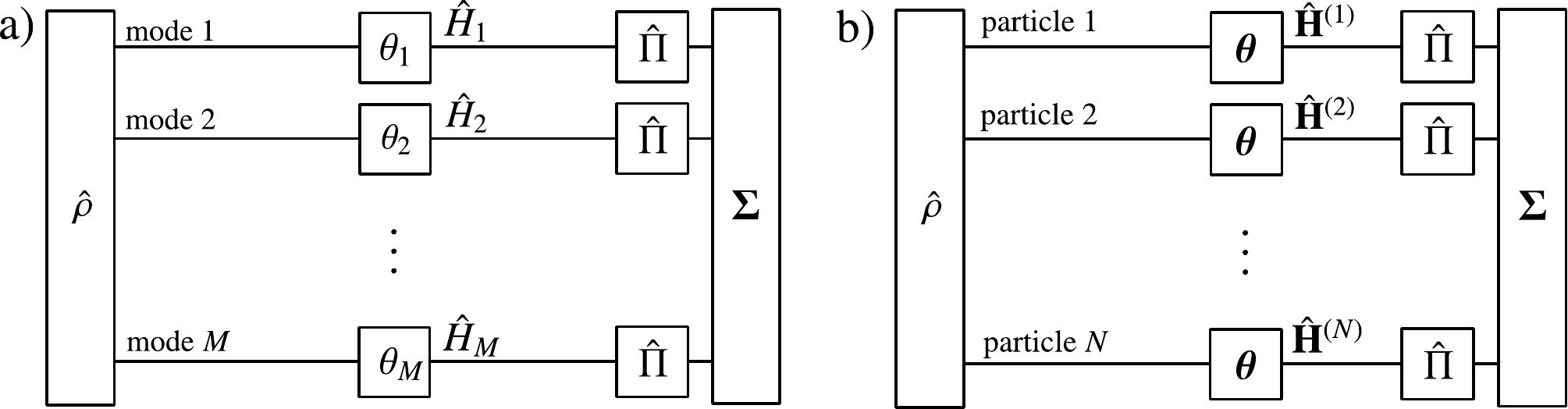}
\caption{The multiparameter estimation with $M$ parameter-encoding modes and $N$ particles can be represented either in the mode (a) or in the particle picture (b). In both cases, the evolution is local. The initial state $\hat{\rho}$ passes through the evolution described by Eq.~(\ref{eq:modeevolution}) and local measurements $\hat{\Pi}$ are performed at the end. The building blocks may be further decomposed as is shown in Fig.~\ref{fig:modevspartB}.}
\label{fig:modevspartA}
\end{figure}

\begin{figure*}[tb]
\centering
\includegraphics[width=.9\textwidth]{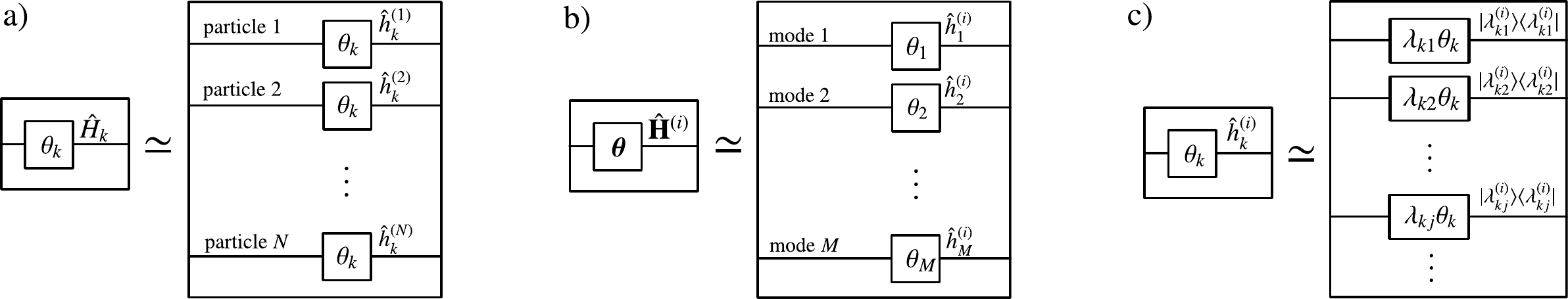}
\caption{The imprinting of a single parameter can be described in terms of all individual particles (a). Every single particle can in principle pass through all the parameter-imprinting modes (b). Each parameter-imprinting mode consists of several sublevels (c). In the main manuscript the case of two sublevels $j=\pm$ with $\lambda_{k+}=-\lambda_{k-}=\frac{1}{2}$ was discussed.}
\label{fig:modevspartB}
\end{figure*}

For a fixed number of $N$ particles and in the absence of particle interactions 
we can represent the multimode interferometer~(\ref{eq:modeevolution}) as a local transformation in the particles. 
We consider the Hilbert space 
$\mathcal{H}=\mathcal{H}^{(1)}\otimes\cdots\otimes\mathcal{H}^{(N)}$, where $\mathcal{H}^{(i)}$ is the Hilbert space of the $i$th particle. 
The Hamiltonian $\hat{H}_k$ can thus be written as 
\begin{align}\label{eq:Hksumi}
\hat{H}_k = \sum_{i=1}^N \hat{h}^{(i)}_k.
\end{align}
where $\hat{h}_k^{(i)}$ is the single particle Hamiltonian on mode $k$, see Fig.~\ref{fig:modevspartB} a).
Hence, in the particle representation, the evolution is described by
\begin{align}\label{eq:particleevolution}
U(\boldsymbol{\theta})=\exp\left(-i\sum_{i=1}^N\hat{\mathbf{H}}^{(i)}\cdot\boldsymbol{\theta}\right),
\end{align}
where $\hat{\mathbf{H}}^{(i)}=(\hat{h}^{(i)}_1,\dots,\hat{h}^{(i)}_M)$ are the local Hamiltonians~(\ref{eq:hik1q}) of particles $i$ in the modes $1,\dots,M$ (Fig.~\ref{fig:modevspartA} b).
It is useful to consider the spectral decomposition of the single-particle Hamiltonian $\hat{h}_k^{(i)}$ (Fig.~\ref{fig:modevspartB} c),  
\begin{align}\label{eq:hik1q}
\hat{h}_k^{(i)}=\sum_j \lambda_{kj}^{(i)} |\lambda^{(i)}_{kj}\rangle\langle \lambda^{(i)}_{kj}|,
\end{align}
where $\lambda_{kj}^{(i)}$  and $|\lambda^{(i)}_{kj}\rangle$ are eigenvalues and corresponding eigenvectors of $\hat{h}_k^{(i)}$, which obey $\langle \lambda^{(i)}_{kj}|\lambda^{(i')}_{k'j'}\rangle=\delta_{ii'}\delta_{kk'}\delta_{jj'}$, and the completeness relation reads
\begin{align}\label{eq:comp}
\quad \sum_{kj} |\lambda^{(i)}_{kj}\rangle\langle \lambda^{(i)}_{kj}|=\mathbb{I}^{(i)},
\end{align}
where $\mathbb{I}^{(i)}$ is the identity on $\mathcal{H}^{(i)}$.

\subsection{Properties of the quantum Fisher and covariance matrices}
\label{Properties}
In the following we analyze some basic properties of the quantum Fisher matrix. First of all, let us notice that we can rewrite the phase imprint transformation~(\ref{eq:modeevolution}) as $\exp\left(-i\hat{\mathbf{H}}\cdot\boldsymbol{\theta}\right)=\exp\left(-i \theta_0 \hat{\mathbf{H}}\cdot \mathbf{n}\right)$, where $\mathbf{n}\in\mathbb{R}^M$ is a vector of real coefficients and $\theta_0$ is a scalar parameter such that $\theta_k = \theta_0 n_k$ for all $k$. The sensitivity of the estimation of the parameter $\theta_0$ is determined by the single-parameter quantum Cram\'er-Rao bound
\begin{align}
(\Delta \theta_{0})^2
\geq \frac{1}{ F_Q[\hat{\rho},\hat{\mathbf{H}}\cdot\mathbf{n}] }. \notag
\end{align}
The quantum Fisher information $F_Q[\hat{\rho},\hat{\mathbf{H}}\cdot\mathbf{n}]$ is related to the quantum Fisher matrix by
\begin{align} \label{eq:singleQFM}
F_Q[\hat{\rho},\hat{\mathbf{H}}\cdot\mathbf{n}]=\mathbf{n}^T\mathbf{F}_Q[\hat{\rho},\hat{\mathbf{H}}]\mathbf{n}.
\end{align}
This can be demonstrated explicitly using the expression of the quantum Fisher in terms of the spectral decomposition of 
$\hat{\rho} = \sum_k p_k \vert k \rangle \langle k \vert$~\cite{BraunsteinPRL1994, Varenna}:
\begin{align}
F_Q[\hat{\rho},\hat{\mathbf{H}}\cdot\mathbf{n}] 
& = 2 \sum_{k,k'}  \frac{(p_k - p_{k'})^2}{p_k + p_{k'}} \vert \langle k \vert \hat{\mathbf{H}}\cdot\mathbf{n} \vert k' \rangle \vert^2 \notag\\
& = 2 \sum_{k,k'}  \frac{(p_k - p_{k'})^2}{p_k + p_{k'}} \bigg\vert \sum_{l=1}^M \langle k \vert  \hat{H}_l  \vert k' \rangle n_l \bigg\vert^2 \notag\\ 
& = 2 \sum_{k,k'}  \frac{(p_k - p_{k'})^2}{p_k + p_{k'}} \left(\sum_{l,l'=1}^M \langle k \vert  \hat{H}_l  \vert k' \rangle\langle k' \vert  \hat{H}_l  \vert k \rangle n_ln_l' \right) \notag\\ 
 & = \mathbf{n}^T\mathbf{F}_Q[\hat{\rho},\hat{\mathbf{H}}]\mathbf{n}. \notag
\end{align}
This implies that the quantum Fisher information matrix shares mathematical properties of the single-parameter quantum Fisher information. We will show this explicitly in the following.

\subsubsection{Convexity of the quantum Fisher matrix}\label{sec:QFMconvex}
Let us consider a convex linear combination of quantum states 
$\hat{\rho}=\sum_{\gamma}p_{\gamma}\hat{\rho}_{\gamma}$. 
From the convexity of the single-parameter quantum Fisher information \cite{Varenna}, $F_Q\left[\sum_{\gamma}p_{\gamma}\hat{\rho}_{\gamma},\hat{\mathbf{H}}\cdot\mathbf{n}\right] 
\leq \sum_{\gamma}p_{\gamma}F_Q\left[\hat{\rho}_{\gamma},\hat{\mathbf{H}}\cdot\mathbf{n}\right]$, we directly obtain that the quantum Fisher matrix is convex too:
\begin{align}\label{eq:QFMconvex}
\mathbf{F}_Q\Big[\sum_{\gamma}p_{\gamma}\hat{\rho}_{\gamma},\hat{\mathbf{H}} \Big] 
\leq \sum_{\gamma}p_{\gamma}\mathbf{F}_Q[\hat{\rho}_{\gamma},\hat{\mathbf{H}}].
\end{align}
This follows since
\begin{align}
\mathbf{n}^T\mathbf{F}_Q\Big[\sum_{\gamma}p_{\gamma}\hat{\rho}_{\gamma},\hat{\mathbf{H}}\Big]\mathbf{n}&=F_Q \Big[ \sum_{\gamma}p_{\gamma}\hat{\rho}_{\gamma},\hat{\mathbf{H}}\cdot\mathbf{n} \Big]\notag\\
&\leq \sum_{\gamma}p_{\gamma}F_Q[\hat{\rho}_{\gamma},\hat{\mathbf{H}}\cdot\mathbf{n}]\notag\\
&= \mathbf{n}^T\left(\sum_{\gamma}p_{\gamma}\mathbf{F}_Q[\hat{\rho}_{\gamma},\hat{\mathbf{H}}]\right)\mathbf{n}, \notag
\end{align}
holds for all $\mathbf{n}$.

\subsubsection{Additivity of the quantum Fisher matrix}
The quantum Fisher information is additive under product states for local evolutions. In the particle representation, for $\hat{\rho}=\hat{\rho}^{(1)}\otimes\cdots\otimes\hat{\rho}^{(N)}$ and $\hat{\mathbf{H}}\cdot\mathbf{n}=\sum_{i=1}^N\hat{\mathbf{H}}^{(i)}\cdot\mathbf{n}$, we have \cite{Varenna} $F_Q\left[\hat{\rho}^{(1)}\otimes\cdots\otimes\hat{\rho}^{(N)},\sum_{i=1}^N\hat{\mathbf{H}}^{(i)}\cdot\mathbf{n}\right]=\sum_{i=1}^NF_Q[\hat{\rho}^{(i)},\hat{\mathbf{H}}^{(i)}\cdot\mathbf{n}]$. This implies the additivity of the quantum Fisher matrix:
\begin{align}\label{eq:QFMaddP}
\mathbf{F}_Q\left[\hat{\rho}^{(1)}\otimes\cdots\otimes\hat{\rho}^{(N)},\sum_{i=1}^N\hat{\mathbf{H}}^{(i)}\right]= \sum_{i=1}^N\mathbf{F}_Q[\hat{\rho}^{(i)},\hat{\mathbf{H}}^{(i)}].
\end{align}
Again, this follows since
\begin{align}
\mathbf{n}^T\mathbf{F}_Q\left[\hat{\rho}^{(1)}\otimes\cdots\otimes\hat{\rho}^{(N)},\hat{\mathbf{H}}\right]\mathbf{n}&=F_Q[\hat{\rho}^{(1)}\otimes\cdots\otimes\hat{\rho}^{(N)},\hat{\mathbf{H}}\cdot\mathbf{n}]\notag\\
&=\sum_{i=1}^NF_Q[\hat{\rho}^{(i)},\hat{\mathbf{H}}^{(i)}\cdot\mathbf{n}]\notag\\
&=\mathbf{n}^T\left(\sum_{i=1}^N\mathbf{F}_Q[\hat{\rho}^{(i)},\hat{\mathbf{H}}^{(i)}]\right)\mathbf{n} \notag
\end{align}
holds for all $\mathbf{n}$.

An analogous result holds also in the mode representation
for Hamiltonians $\hat{\mathbf{H}}\cdot\mathbf{n}=\sum_{k=1}^M\hat{H}_kn_k$ and mode-product states
$\hat{\rho}_1\otimes\cdots\otimes\hat{\rho}_M$, where $\hat{H}_k$ and $\hat{\rho}_k$ act on the Hilbert space $\mathcal{H}_k$ of mode $k$. Additivity leads to a diagonal Fisher matrix for mode-product states:
\begin{align}\label{eq:QFMaddM}
\mathbf{F}_Q[\hat{\rho}_1\otimes\cdots\otimes\hat{\rho}_M,\hat{\mathbf{H}}]=\begin{pmatrix}F_Q[\hat{\rho}_1,\hat{H}_1]& 0 & \cdots & 0\\
\vdots & &\ddots & \vdots\\
0 & \cdots & 0 &F_Q[\hat{\rho}_M,\hat{H}_M]\end{pmatrix}.
\end{align}
This follows from $\mathbf{n}^T\mathbf{F}_Q[\hat{\rho}_1\otimes\cdots\otimes\hat{\rho}_M,\hat{\mathbf{H}}]\mathbf{n}=F_Q[\hat{\rho}_1\otimes\cdots\otimes\hat{\rho}_M,\hat{\mathbf{H}}\cdot\mathbf{n}]=\sum_{k=1}^MF_Q[\hat{\rho}_k,\hat{H}_kn_k]=\mathbf{n}^T\mathrm{diag}(F_Q[\hat{\rho}_1,\hat{H}_1],\dots,F_Q[\hat{\rho}_M,\hat{H}_M])\mathbf{n}$, which holds for all $\mathbf{n}$.

\subsubsection{Relation between quantum Fisher matrix and the covariance matrix}
For pure states the quantum Fisher information coincides with four times the variance~\cite{BraunsteinPRL1994}, i.e., $F_Q[|\Psi\rangle,\hat{\mathbf{H}}\cdot\mathbf{n}]=4\Delta \left(\hat{\mathbf{H}}\cdot\mathbf{n}\right)^2_{|\Psi\rangle}$. Similarly, for pure states the quantum Fisher matrix coincides with four times the covariance matrix, i.e.,
\begin{align}\label{eq:QFMpure}
\mathbf{F}_Q[|\Psi\rangle,\hat{\mathbf{H}}]=4\boldsymbol{\Gamma}[|\Psi\rangle,\hat{\mathbf{H}}],
\end{align}
where $\left(\boldsymbol{\Gamma}[\hat{\rho},\hat{\mathbf{H}}]\right)_{kl}=\mathrm{Cov}(\hat{H}_k,\hat{H}_l)_{\hat{\rho}}$, with
\begin{align} 
\mathrm{Cov}(\hat{H}_k,\hat{H}_l)_{\hat{\rho}}
=\frac{1}{2}\left(\langle\hat{H}_k\hat{H}_l\rangle_{\hat{\rho}}+\langle\hat{H}_l\hat{H}_k\rangle_{\hat{\rho}}\right)-\langle\hat{H}_k\rangle_{\hat{\rho}}\langle\hat{H}_l\rangle_{\hat{\rho}}.\notag
\end{align}
This follows from $\mathbf{n}^T\mathbf{F}_Q[|\Psi\rangle,\hat{\mathbf{H}}]\mathbf{n}=F_Q[|\Psi\rangle,\hat{\mathbf{H}}\cdot\mathbf{n}]=4\Delta (\hat{\mathbf{H}}\cdot\mathbf{n})^2_{|\Psi\rangle}=4 \mathbf{n}^T\boldsymbol{\Gamma}[|\Psi\rangle,\hat{\mathbf{H}}]\mathbf{n}$, that holds for all $\mathbf{n}$. Here we used the bilinearity property
\begin{align}\label{eq:bilcov}
\mathrm{Cov}\left(\sum_{k}n_{k}\hat{H}_k,\sum_{l}n_{l}\hat{H}_l\right)_{\hat{\rho}}&=\sum_{kl}n_{k}n_{l}\mathrm{Cov}(\hat{H}_k,\hat{H}_l)_{\hat{\rho}},
\end{align}
which implies
\begin{align} \label{eq:covvar}
\Delta \left(\hat{\mathbf{H}}\cdot\mathbf{n}\right)^2_{\hat{\rho}}=\mathbf{n}^T\boldsymbol{\Gamma}[\hat{\rho},\hat{\mathbf{H}}]\mathbf{n}.
\end{align}

For mixed states, the covariance yields an upper bound on the quantum Fisher information \cite{BraunsteinPRL1994}, i.e., $F_Q[\hat{\rho},\hat{\mathbf{H}}\cdot\mathbf{n}]\leq 4\Delta (\hat{\mathbf{H}}\cdot\mathbf{n})^2_{\hat{\rho}}$, for arbitrary $\hat{\rho}$. Analogously, this implies the matrix inequality
\begin{align}\label{eq:QFMvar}
\mathbf{F}_Q[\hat{\rho},\hat{\mathbf{H}}] \leq 4 \boldsymbol{\Gamma}[\hat{\rho},\hat{\mathbf{H}}],
\end{align}
for arbitrary $\hat{\rho}$.

\subsubsection{Concavity of the covariance matrix}
Notice that, in contrast to the quantum Fisher matrix, which is convex [Eq.~(\ref{eq:QFMconvex})], the covariance matrix is concave:
\begin{align}\label{eq:covconcave}
\boldsymbol{\Gamma}\Big[\sum_{\gamma}p_{\gamma}\hat{\rho}_\gamma,\hat{\mathbf{H}}\Big]
\geq \sum_{\gamma}p_{\gamma}\boldsymbol{\Gamma}[\hat{\rho}_\gamma,\hat{\mathbf{H}}].
\end{align}
To see this recall that for a linear combination of quantum states $\hat{\rho}=\sum_{\gamma}p_{\gamma}\hat{\rho}_\gamma$, the variance is concave:
\begin{align}\label{eq:normalvarconcave}
\Delta \left(\hat{\mathbf{H}}\cdot\mathbf{n}\right)^2_{\sum_{\gamma}p_{\gamma}\hat{\rho}_\gamma }\geq \sum_{\gamma}p_{\gamma}\Delta \left(\hat{\mathbf{H}}\cdot\mathbf{n}\right)^2_{\hat{\rho}_\gamma}.
\end{align}
Using Eq.~(\ref{eq:covvar}), this inequality becomes $\mathbf{n}^T\boldsymbol{\Gamma}[\sum_{\gamma}p_{\gamma}\hat{\rho}_\gamma,\hat{\mathbf{H}}]\mathbf{n}\geq \mathbf{n}^T\left(\sum_{\gamma}p_{\gamma}\boldsymbol{\Gamma}[\hat{\rho}_\gamma,\hat{\mathbf{H}}]\right)\mathbf{n}$, which holds for arbitrary $\mathbf{n}$ and therefore implies Eq.~(\ref{eq:covconcave}).

\subsubsection{Upper bound for the covariance matrix}
The covariance matrix is upper bounded by:
\begin{align}\label{eq:upperbounddiag}
\boldsymbol{\Gamma}[\hat{\rho},\hat{\mathbf{H}}]\leq\tilde{\boldsymbol{\Gamma}}[\hat{\rho},\hat{\mathbf{H}}],
\end{align}
where
\begin{align}
\left(\tilde{\boldsymbol{\Gamma}}[\hat{\rho},\hat{\mathbf{H}}]\right)_{ij}=\frac{1}{2}\left(\langle\hat{H}_k\hat{H}_l\rangle_{\hat{\rho}}+\langle\hat{H}_l\hat{H}_k\rangle_{\hat{\rho}}\right) \notag
\end{align}
contains only the fluctuations. 
This can be immediately demonstrated noticing that 
\begin{align}
\tilde{\boldsymbol{\Gamma}}[\hat{\rho},\hat{\mathbf{H}}]-\boldsymbol{\Gamma}[\hat{\rho},\hat{\mathbf{H}}]=\begin{pmatrix} H_1^2 & H_1H_2 & \dots & H_1H_M\\
H_1H_2 & H_2^2 & \cdots & H_2H_M\\
\vdots & & \ddots & \vdots\\
H_1H_M & H_2H_M & \cdots  & H_M^2
\end{pmatrix}=\mathbf{H}\mathbf{H}^T, \notag
\end{align}
where $\mathbf{H}=(H_1,\dots,H_M)$ is the vector of mean values $H_k=\langle\hat{H}_k\rangle_{\hat{\rho}}$. The above matrix is of rank one with eigenvalue $\mathbf{H}^T\mathbf{H}=\sum_{k=1}^NH_k^2$, which is clearly non-negative. The statement holds for all $\hat{\rho}$ and the bound is saturated by states with the property $H_k=0$ for all $k$. 

\section{Bounds for particle-separable states}
\label{BoundsPartSep}
In this section we derive upper bounds on the quantum Fisher matrix for particle separable states,
\begin{align}
\hat{\rho}_{\mathrm{p-sep}}=\sum_{\gamma}p_{\gamma}\hat{\rho}^{(1)}_\gamma \otimes\cdots\otimes\hat{\rho}^{(N)}_\gamma,\notag
\end{align}
where $p_{\gamma}$ is a probability distribution and $\hat{\rho}^{(i)}_\gamma$ are single-particle density matrices on 
$\mathcal{H}^{(i)}$ for the particles $i=1,\dots,N$.

\subsection{State-dependent bounds}
Recall from Eq.~(\ref{eq:particleevolution}) that the phase encoding is local in the particles, i.e., $\hat{\mathbf{H}}=\sum_{i=1}^N\hat{\mathbf{H}}^{(i)}$. We first use the convexity~(\ref{eq:QFMconvex}) and additivity~(\ref{eq:QFMaddP}) properties of the quantum Fisher matrix. Then we use the upper bound~(\ref{eq:QFMvar}) and the concavity~(\ref{eq:covconcave}) of the variance. We obtain
\begin{align}
\mathbf{F}_Q[\hat{\rho}_{\mathrm{p-sep}},\hat{\mathbf{H}}]&\stackrel{(\ref{eq:QFMconvex})}{\leq} 
\sum_{\gamma}p_{\gamma}\mathbf{F}_Q[\hat{\rho}^{(1)}_\gamma \otimes\cdots\otimes\hat{\rho}^{(N)}_\gamma,\hat{\mathbf{H}}]\notag\\
&\stackrel{(\ref{eq:QFMaddP})}{=}\sum_{\gamma}p_{\gamma}\sum_{i=1}^N\mathbf{F}_Q[\hat{\rho}^{(i)}_\gamma,\hat{\mathbf{H}}^{(i)}]\notag\\
&\stackrel{(\ref{eq:QFMvar})}{\leq} 4\sum_{\gamma}p_{\gamma}\sum_{i=1}^N \boldsymbol{\Gamma}[\hat{\rho}^{(i)}_\gamma,\hat{\mathbf{H}}^{(i)}]\notag\\
&\stackrel{(\ref{eq:covconcave})}{\leq} 4\sum_{i=1}^N \boldsymbol{\Gamma}[\hat{\rho}^{(i)},\hat{\mathbf{H}}^{(i)}], \notag
\end{align}
where $\hat{\rho}^{(i)}=\sum_{\gamma}p_{\gamma}\hat{\rho}^{(i)}_\gamma$ is the reduced density matrix of particle $i$. 
We thus have
\begin{align}\label{eq:pseplimitmtx}
\mathbf{F}_Q[\hat{\rho}_{\mathrm{p-sep}}]\leq 4 \boldsymbol{\Gamma}[\hat{\rho}^{(1)}\otimes\cdots\otimes\hat{\rho}^{(N)},\hat{\mathbf{H}}],
\end{align}
with
\begin{align}\label{eq:npartcov}
\boldsymbol{\Gamma}[\hat{\rho}^{(1)}\otimes\cdots\otimes\hat{\rho}^{(N)},\hat{\mathbf{H}}]=\sum_{i=1}^N\boldsymbol{\Gamma}[\hat{\rho}^{(i)},\hat{\mathbf{H}}^{(i)}]
\end{align}
being the covariance matrix of the product state $\hat{\rho}^{(1)} \otimes\cdots\otimes\hat{\rho}^{(N)}$ 
of reduced density matrices \cite{GessnerPRA2016}.

\subsection{Multiparameter shot-noise limit}
To find the multiparameter shot-noise limit, we maximize $\mathbf{F}_Q[\hat{\rho}_{\mathrm{p-sep}},\hat{\mathbf{H}}]$ over all possible particle separable states $\hat{\rho}_{\mathrm{p-sep}}$. The convexity of the Fisher information allows us to limit the optimization problem to $N$-particle pure product states as these states achieve equality in~(\ref{eq:pseplimitmtx}). We thus have 
\begin{align}
\mathbf{F}_{\mathrm{SN}}:=&\max_{\rho_{\mathrm{p-sep}}}\mathbf{F}_Q[\hat{\rho}_{\mathrm{p-sep}},\hat{\mathbf{H}}]\notag\\=&\max_{|\Psi^{(1)}\rangle\otimes\cdots\otimes|\Psi^{(N)}\rangle}4\boldsymbol{\Gamma}[|\Psi^{(1)}\rangle\otimes\cdots\otimes|\Psi^{(N)}\rangle,\hat{\mathbf{H}}].\notag
\end{align}
Recalling $\hat{\mathbf{H}}^{(i)}=(\hat{h}_1^{(i)},\dots,\hat{h}_M^{(i)})$, the elements of the covariance matrix for the single-particle pure state $|\Psi^{(i)}\rangle\in\mathcal{H}^{(i)}$ are given by $\mathrm{Cov}(\hat{h}^{(i)}_k,\hat{h}^{(i)}_l)_{|\Psi^{(i)}\rangle}=
\delta_{kl}\langle \hat{h}^{(i)2}_k \rangle_{|\Psi^{(i)}\rangle}- \langle \hat{h}^{(i)}_k \rangle_{|\Psi^{(i)}\rangle}\langle \hat{h}^{(i)}_l \rangle_{|\Psi^{(i)}\rangle}$, where we used that $\hat{h}^{(i)}_k\hat{h}^{(i)}_l=\delta_{kl}\hat{h}^{(i)2}_k$ in the single-particle subspace, as can be verified from Eq.~(\ref{eq:hik1q}). 
We can therefore express the single-particle covariance matrix as $\boldsymbol{\Gamma}[|\Psi^{(i)}\rangle,\hat{\mathbf{H}}^{(i)}]=\mathbf{D}^{(i)}-\mathbf{h}^{(i)}\mathbf{h}^{(i)T}$, where $\mathbf{D}^{(i)}=\mathrm{diag}(d^{(i)}_1,\dots,d^{(i)}_M)$, with $d^{(i)}_k=\langle\Psi^{(i)}|\hat{h}^{(i)2}_k|\Psi^{(i)}\rangle$ is a 
diagonal $M\times M$ matrix containing the fluctuations and $\mathbf{h}^{(i)}=(h^{(i)}_1,\dots,h^{(i)}_M)$ is a vector of mean values $h^{(i)}_k=\langle\Psi^{(i)}|\hat{h}^{(i)}_k|\Psi^{(i)}\rangle$. The full $N$-particle covariance matrix, Eq.~(\ref{eq:npartcov}), is obtained by summing over all single-particle terms:
\begin{align}\label{eq:purepartprodstatecov}
\boldsymbol{\Gamma}[|\Psi^{(1)}\rangle\otimes\cdots\otimes|\Psi^{(N)}\rangle,\hat{\mathbf{H}}]&=\sum_{i=1}^N\boldsymbol{\Gamma}[|\Psi^{(i)}\rangle,\hat{\mathbf{H}}^{(i)}]\notag\\&=\sum_{i=1}^N\mathbf{D}^{(i)}-\sum_{i=1}^N\mathbf{h}^{(i)}\mathbf{h}^{(i)T}.
\end{align}
By virtue of Eq.~(\ref{eq:upperbounddiag}) we have $\boldsymbol{\Gamma}[|\Psi^{(i)}\rangle,\hat{\mathbf{H}}^{(i)}]\leq\mathbf{D}^{(i)}$.
Maximizing over all single-particle pure states, we obtain $\mathbf{F}_{\mathrm{SN}} \leq 4\sum_{i=1}^N\max_{|\Psi^{(i)}\rangle}\mathbf{D}^{(i)}$. Furthermore, since $\mathbf{D}^{(i)}$ is diagonal matrix, the maximization can be carried out element-wise. 
To accomplish this we consider the spectral decomposition of $\hat{h}^{(i)}$, Eq.~(\ref{eq:hik1q}), and obtain 
\begin{align}\label{eq:maxfordi}
d^{(i)}_k=\sum_{j}\lambda_{kj}^2|\langle\Psi^{(i)}|\lambda^{(i)}_{kj}\rangle|^2 \leq \lambda_{k\max}^2 p^{(i)}_k,
\end{align}
where $\lambda_{k\max}=\max_j\{|\lambda_{kj}|\}$ is the maximum eigenvalue of $|\hat{h}^{(i)}_k|$, and
\begin{align}\label{eq:pik}
p^{(i)}_k=\sum_{j}|\langle\Psi^{(i)}|\lambda^{(i)}_{kj}\rangle|^2
\end{align}
is the probability to find particle $i$ in mode $k$ (with $\sum_{k=1}^Mp_k^{(i)}=1$). 
Performing the sum over all $N$ particles yields
\begin{equation}
\sum_{i=1}^Nd^{(i)}_k \leq \lambda^2_{k\max} \sum_{i=1}^N p_k^{(i)}  = \lambda^2_{k\max} \langle \hat{N}_k\rangle, \notag
\end{equation}
where $\langle \hat{N}_k\rangle = \sum_{i=1}^N p_k^{(i)}$ is the average number of particles in the mode $k$.
Hence, we obtain
\begin{align}\label{eq:psebound}
\mathbf{F}_{\mathrm{SN}}\leq 4\begin{pmatrix}\langle\hat{N}_1\rangle \lambda_{1\max}^2& 0 & \cdots & 0\\
\vdots & &\ddots & \vdots\\
0 & \cdots & 0 &\langle\hat{N}_M\rangle\lambda_{M\max}^2\end{pmatrix}.
\end{align}
This upper bound is valid for arbitrary Hamiltonians and quantum states. 
It is saturated if and only if
\begin{align}
h_k^{(i)}&=0\label{eq:condspc1}
\end{align}
and
\begin{align}
d^{(i)}_k &= \lambda^2_{k\max}p_k^{(i)}. \label{eq:condspc2}
\end{align}
Both conditions can be satisfied only for single-particle Hamiltonians with the
following property:
\begin{align}\label{eq:property}
\lambda_{k+}=-\lambda_{k-}\quad\mathrm{for}\:\mathrm{all}\quad k=1,\dots,M,
\end{align}
where $\lambda_{k+}=\max_j\lambda_{kj}$ and $\lambda_{k-}=\min_j\lambda_{kj}$ denote the largest and smallest eigenvalue of $\hat{h}^{(i)}_k$, respectively. Notice that if the property~(\ref{eq:property}) is valid, we may write $4\lambda_{k\max}^2=(\lambda_{k+}-\lambda_{k-})^2$, which in the single-mode case reduces to the well-known form of the shot-noise limit~\cite{GiovannettiPRL2006}. Equation~(\ref{eq:property}) ensures that there exists a single-particle quantum state that reaches the maximum~(\ref{eq:maxfordi}) for $d^{(i)}_k$ while yielding $h^{(i)}_k=0$ at the same time. Physically, this condition can be interpreted as follows. Since the phase shift $\theta_k$ can be detected with the highest sensitivity if it is imprinted with the largest possible $|\lambda_{kj}|$, we may restrict our treatment to the extremal levels. Condition~(\ref{eq:property}) now imposes that the phase shift can be acquired as a relative, balanced phase shift between the two extremal levels, i.e., both levels contribute with equal weight.

\subsection{Optimal particle-separable states}
Here we discuss the quantum states which saturate the upper sensitivity bound~(\ref{eq:psebound}) for Hamiltonians with the property~(\ref{eq:property}). 
Using the completeness relation (\ref{eq:comp}), a pure single-particle state $|\Psi^{(i)}\rangle\in\mathcal{H}^{(i)}$ can be expanded as
\begin{align}\label{eq:spstate}
|\Psi^{(i)}\rangle=\sum_{kj}c^{(i)}_{kj}|\lambda^{(i)}_{kj}\rangle, 
\end{align}
where $c^{(i)}_{kj}=\langle\lambda^{(i)}_{kj}|\Psi^{(i)}\rangle$. The covariance matrix is entirely determined by the joint probabilities $p_{kj}^{(i)}=|c^{(i)}_{kj}|^2$ to find particle $i$ in sublevel $j$ of mode $k$:
\begin{align}
h_k^{(i)}=\sum_j\lambda_{kj}p^{(i)}_{kj}, \qquad
d_k^{(i)}=\sum_j\lambda_{kj}^2p^{(i)}_{kj}.\label{eq:dik}
\end{align}
We may decompose the joint probability as
\begin{align}\label{eq:pijkpik}
p^{(i)}_{kj}=p^{(i)}_{j|k}p^{(i)}_k,
\end{align}
where $p^{(i)}_k$ was defined in Eq.~(\ref{eq:pik}) and $p^{(i)}_{j|k}$ is the conditional probability to find the particle $i$ in state $j$, given that the particle $i$ is in some state of mode $k$. This distribution satisfies the normalization condition $\sum_jp^{(i)}_{j|k}=1$.

Optimal single-particle states are those that maximize $d_k^{(i)} = p^{(i)}_k \sum_j \lambda_{kj}^2 p^{(i)}_{j|k}$, with $h_k^{(i)}=0$. 
The maximization is thus independent of $p^{(i)}_k$ and is obtained for 
\begin{align}\label{eq:optimalcond}
p_{+|k}^{(i)}=p_{-|k}^{(i)}=\frac{1}{2},
\end{align}
corresponding to an equal distribution among the maximum and minimum values of $\lambda_{kj}$.
If Eq.~(\ref{eq:property}) holds, we see immediately that states with this property fulfill both conditions~(\ref{eq:condspc1}) and~(\ref{eq:condspc2}), and thus saturate the upper sensitivity bound~(\ref{eq:psebound}) for particle-separable states. Notice that Eq.~(\ref{eq:optimalcond}) only determines the conditional probabilities $p^{(i)}_{j|k}$, i.e., the quantum state within the modes $k$, but does not depend on the distribution of particles among modes, i.e., the $p^{(i)}_k$. If $\langle \hat{N}_k\rangle$ is integer, we may therefore saturate the shot-noise limit by sending each particle in a specific single mode $k_i$, where it realizes a superposition of largest and smallest eigenvalue of the kind
\begin{align}
|\Psi^{(i)}\rangle=\frac{1}{\sqrt{2}}(|\lambda^{(i)}_{k_i+}\rangle+|\lambda^{(i)}_{k_i-}\rangle).\notag
\end{align}
This corresponds to choosing the $p^{(i)}_k=\delta_{kk_i}$ such that a total number of $\langle \hat{N}_k\rangle$ particles enter the mode $k$, i.e., $\sum_{i=1}^Np^{(i)}_k=\sum_{i=1}^N\delta_{kk_i}=\langle \hat{N}_k\rangle$. 
To summarize, optimal particle-separable states are characterized by an optimal distribution~(\ref{eq:optimalcond}) of particles within each mode $k$, 
but the sensitivity is independent of the delocalization of the particles over the modes. 
This is only true for states with $h^{(i)}_k=0$, as we will see in the next section.

\subsection{Optimizing the distribution of individual particles among modes}\label{sec:optpi}
The upper bound~(\ref{eq:psebound}) is independent of the distribution of particles among modes, i.e., of the $p^{(i)}_k$, 
but can only be saturated for multimode interferometers with balanced local evolutions, described by~(\ref{eq:property}). 
In this section, we release the condition~(\ref{eq:property}) and maximize the covariance matrix~(\ref{eq:purepartprodstatecov})
for arbitrary Hamiltonians.
In other words, we search for the optimal choice of $p^{(i)}_k$ that lead to the smallest 
covariance matrix in Eq.~(\ref{eq:purepartprodstatecov}) for any given fixed choice of the $p_{j|k}^{(i)}$.

We begin by focusing on the first part in Eq.~(\ref{eq:purepartprodstatecov}), i.e., the fluctuations. 
We assume the $p_{j|k}$ to be independent of $i$ since all particles experience the same evolution and therefore 
the optimal quantum state within each mode is independent of the particle label.
The matrix $\sum_{i=1}^N\mathbf{D}^{(i)}$ is diagonal with elements 
\begin{align}\label{eq:spfluc}
\sum_{i=1}^Nd^{(i)}_k=\sum_{i=1}^N\sum_{j}\lambda^2_{kj}p_{j|k}p^{(i)}_k=\langle \hat{N}_k\rangle\sum_{j}\lambda^2_{kj}p_{j|k}.
\end{align}
This is independent of our choice for the individual $p^{(i)}_k$ as it only depends on their sum which is always fixed by the number of particles in mode $k$. 
For the maximization of the covariance matrix, we can therefore focus entirely on the second part of  Eq.~(\ref{eq:purepartprodstatecov}), i.e., $\sum_{i=1}^N\mathbf{h}^{(i)}\mathbf{h}^{(i)T}$. 
The following result reveals the sensitivities of different strategies for the distribution of particles among the modes by varying the coefficients $p^{(i)}_k$ for any fixed choice of the $p_{j|k}$.

The covariance matrix of arbitrary pure product states $|\Psi^{(1)}\rangle\otimes\cdots\otimes|\Psi^{(N)}\rangle$ with fixed $p_{j|k}$ is bounded from below and above by:
\begin{align}\label{eq:optspcov}
\boldsymbol{\Gamma}_{\mathrm{MsPs}}\leq \boldsymbol{\Gamma}[|\Psi^{(1)}\rangle\otimes\cdots\otimes|\Psi^{(N)}\rangle,\hat{\mathbf{H}}]\leq \boldsymbol{\Gamma}_{\mathrm{MePs}}.
\end{align} 
Here, $\boldsymbol{\Gamma}_{\mathrm{MePs}}$ is the covariance matrix~(\ref{eq:purepartprodstatecov}) obtained by fully delocalized single-particle states $|\Psi^{(i)}\rangle$, as in Eq.~(\ref{eq:spstate}) with uniform $p^{(i)}_k=\frac{\langle \hat{N}_k\rangle}{N}$ for all $i$. Moreover, $\boldsymbol{\Gamma}_{\mathrm{MsPs}}$ is the covariance matrix~(\ref{eq:purepartprodstatecov}) obtained from fully localized single-particle states, i.e., by choosing $p_k^{(i)}=\delta_{kk_i}$  such that $\sum_{i=1}^N\delta_{kk_i}=\langle \hat{N}_k\rangle$, which can only produce integer values of $\langle \hat{N}_k\rangle$. 
The theorem states that the delocalization of particles in the $k$ modes (or in other words mode entanglement) leads, in general, to a higher sensitivity bound. 
For states with $h^{(i)}_k=0$ for all $k$, however, there is no difference between the different strategies and all inequalities in~(\ref{eq:optspcov}) become equalities. 
Notice that this statement only discusses the distribution of the particles among the modes, i.e., the $p_k^{(i)}$. 
The the quantum states that the particles assume within the modes, i.e., the $p_{j|k}$, are chosen the same on all sides of the inequalities~(\ref{eq:optspcov}).

To prove the upper bound of the statement~(\ref{eq:optspcov}), we must show that
\begin{align}\label{eq:zzoptspcov}
\mathbf{v}^T(\boldsymbol{\Gamma}_{\mathrm{MePs}}-\boldsymbol{\Gamma}[|\Psi^{(1)}\rangle\otimes\cdots\otimes|\Psi^{(N)}\rangle,\hat{\mathbf{H}}])\mathbf{v}\geq 0
\end{align}
for arbitrary vectors $\mathbf{v}=(v_1,\dots,v_M)$. Since the fluctuation part [i.e., $\sum_{i=1}^N\mathbf{D}^{(i)}$ in Eq.~(\ref{eq:purepartprodstatecov})] is always independent of the choice of the individual $p_k^{(i)}$ [recall Eq.~(\ref{eq:spfluc})], the difference between the two matrices is given by
\begin{align}
\boldsymbol{\Gamma}_{\mathrm{MePs}}-\boldsymbol{\Gamma}[|\Psi^{(1)}\rangle\otimes\cdots\otimes|\Psi^{(N)}\rangle,\hat{\mathbf{H}}]=\sum_{i=1}^N\mathbf{h}^{(i)}\mathbf{h}^{(i)T}-N\mathbf{h}_{\mathrm{del}}\mathbf{h}_{\mathrm{del}}^T,\notag
\end{align}
where $\mathbf{h}^{(i)}=(\mathfrak{h}_1p^{(i)}_1,\dots,\mathfrak{h}_Mp^{(i)}_M)$ and $\mathbf{h}_{\mathrm{del}}=(\mathfrak{h}_1\langle\hat{N}_1\rangle,\dots,\mathfrak{h}_M\langle\hat{N}_M\rangle)/N$ with $\mathfrak{h}_k=\sum_j\lambda_{kj}p_{j|k}$ are the mean values of the arbitrary and the delocalized state, respectively, according to Eqs.~(\ref{eq:dik}) and~(\ref{eq:pijkpik}). Inequality~(\ref{eq:zzoptspcov}) thus reads
\begin{align}\label{eq:zzoptcoeff}
\sum_{i=1}^N\left(\sum_{k}^Mv_k\mathfrak{h}_kp^{(i)}_k\right)^2\geq\left(\sum_{k=1}^M\frac{v_k\mathfrak{h}_k\langle \hat{N}_k\rangle}{\sqrt{N}}\right)^2.
\end{align}
Let us now recall the Cauchy-Schwarz inequality $\left(\sum_{i=1}^Nf_i^2\right)\left(\sum_{i=1}^Ng_i^2\right)\geq \left(\sum_{i=1}^Nf_ig_i\right)^2$
for arbitrary vectors with real-valued elements $f_i$ and $g_i$ and finite norms. Choosing $f_i=\sum_{k=1}^Mv_k\mathfrak{h}_kp^{(i)}_k$ and $g_i=1/\sqrt{N}$, we obtain $\left(\sum_{i=1}^Ng_i^2\right)=\left(\sum_{i=1}^N\frac{1}{N}\right)=1$ and 
\begin{align}
\sum_{i=1}^Nf_ig_i=\sum_{i=1}^N\sum_{k=1}^M\frac{v_k\mathfrak{h}_kp^{(i)}_k}{\sqrt{N}}=\sum_{k=1}^M\frac{v_k\mathfrak{h}_k\langle \hat{N}_k\rangle}{\sqrt{N}},\notag
\end{align}
since $\sum_{i=1}^Np^{(i)}_k=\langle \hat{N}_k\rangle$. Hence,~(\ref{eq:zzoptcoeff}) follows from the Cauchy-Schwarz inequality and thus~(\ref{eq:zzoptspcov}) holds for arbitrary $\mathbf{v}$ and we have proven the upper bound in the statement~(\ref{eq:optspcov}).

In order to prove the lower bound, we need to show that
\begin{align}
\mathbf{v}^T(\boldsymbol{\Gamma}[|\Psi^{(1)}\rangle\otimes\cdots\otimes|\Psi^{(N)}\rangle,\hat{\mathbf{H}}]-\boldsymbol{\Gamma}_{\mathrm{MsPs}})\mathbf{v}\geq 0,\notag
\end{align}
holds for arbitrary $\mathbf{v}$. This can be expressed in elements as
\begin{align}\label{eq:zzlowerbnd}
\sum_{k,l=1}^Mv_kv_l\mathfrak{h}_k\mathfrak{h}_l\sum_{i=1}^N\delta_{kk_i}\delta_{lk_i}\geq\sum_{i=1}^N\left(\sum_{k=1}^Mv_k\mathfrak{h}_kp^{(i)}_k\right)^2.
\end{align}
Since in the fully localized strategy, a single particle is either localized in mode $k$ or in mode $l$, we obtain that $\sum_{i=1}^N\delta_{kk_i}\delta_{lk_i}=\delta_{kl}\sum_{i=1}^N\delta_{kk_i}=\delta_{kl}\langle \hat{N}_k\rangle=\delta_{kl}\sum_{i=1}^Np_k^{(i)}$. With this we can write the left-hand side of~(\ref{eq:zzlowerbnd}) as $\sum_{i=1}^N\sum_{k=1}^Mv_k^2\mathfrak{h}_k^2p^{(i)}_k$. The inequality 
\begin{align}\label{eq:zzlbnd2}
\sum_{k=1}^Mv_k^2\mathfrak{h}_k^2p^{(i)}_k\geq\left(\sum_{k=1}^Mv_k\mathfrak{h}_kp^{(i)}_k\right)^2
\end{align}
follows from the Cauchy-Schwarz inequality $\left(\sum_{k=1}^Mf_k^2\right)\left(\sum_{k=1}^Mg_k^2\right)\geq \left(\sum_{k=1}^Mf_kg_k\right)^2$ with $f_k=v_k\mathfrak{h}_k\sqrt{p^{(i)}_k}$ and $g_k=\sqrt{p^{(i)}_k}$, which leads to $\sum_{k=1}^Mg_k^2=\sum_{k=1}^Mp^{(i)}_k=1$. Finally, the inequality~(\ref{eq:zzlowerbnd}) follows by summation of~(\ref{eq:zzlbnd2}) over $i$. This concludes the proof for the lower bound in~(\ref{eq:optspcov}).

In the main article we state the equivalent result for the quantum Fisher matrix using the equivalence~(\ref{eq:QFMpure}) for pure states. Clearly, if $\mathfrak{h}_k=0$ for all $k$, the difference between the three matrices~(\ref{eq:optspcov}) vanishes. 

\section{Bounds for mode-separable states}
In this section, we maximize the quantum Fisher matrix for mode-separable states 
\begin{align}
\hat{\rho}_{\mathrm{m-sep}}=\sum_{\gamma}p_{\gamma}\hat{\rho}_{\gamma,1}\otimes\cdots\otimes\hat{\rho}_{\gamma,M},\notag
\end{align}
where $\hat{\rho}_{\gamma,k}$ is an arbitrary density matrix on the Hilbert space $\mathcal{H}_k$ of mode $k=1,\dots,M$.

\subsection{State-dependent bounds}
Using the convexity property~(\ref{eq:QFMconvex}), additivity~(\ref{eq:QFMaddM}), the upper bound~(\ref{eq:QFMvar}), and the concavity of the covariance~(\ref{eq:normalvarconcave}):
\begin{align} 
\mathbf{F}_Q[\hat{\rho}_{\mathrm{m-sep}},\hat{\mathbf{H}}]&\stackrel{(\ref{eq:QFMconvex})}{\leq} \sum_{\gamma}p_{\gamma}\mathbf{F}_Q[\hat{\rho}_{\gamma,1}\otimes\cdots\otimes\hat{\rho}_{\gamma,M},\hat{\mathbf{H}}]\notag\\
&\stackrel{(\ref{eq:QFMaddM})}{\leq} \sum_{\gamma}p_{\gamma} \begin{pmatrix}F_Q[\hat{\rho}_{\gamma,1},\hat{H}_1]& 0 & \cdots & 0\\
\vdots & &\ddots & \vdots\\
0 & \cdots & 0 &F_Q[\hat{\rho}_{\gamma,M},\hat{H}_M]\end{pmatrix}\notag\\
&\stackrel{(\ref{eq:QFMvar})}{\leq} 4\begin{pmatrix}\sum_{\gamma}p_{\gamma}(\Delta\hat{H}_1)^2_{\hat{\rho}_{\gamma,1}}& 0 & \cdots & 0\\
\vdots & &\ddots & \vdots\\
0 & \cdots & 0 &\sum_{\gamma}p_{\gamma}(\Delta\hat{H}_M)^2_{\hat{\rho}_{\gamma,M}}\end{pmatrix}\notag\\
&\stackrel{(\ref{eq:covconcave})}{\leq} 4 \boldsymbol{\Gamma}[\hat{\rho}_1\otimes\cdots\otimes\hat{\rho}_M,\hat{\mathbf{H}}], \notag
\end{align}
where $\hat{\rho}_k=\sum_{\gamma}p_{\gamma}\hat{\rho}_{\gamma,k}$ is the reduced density matrix of mode $k$~\cite{GessnerPRA2016}. The upper bound,
\begin{align}\label{eq:mseplimitmtx}
\mathbf{F}_Q[\hat{\rho}_{\mathrm{m-sep}},\hat{\mathbf{H}}]\leq 4 \boldsymbol{\Gamma}[\hat{\rho}_1\otimes\cdots\otimes\hat{\rho}_M,\hat{\mathbf{H}}]
\end{align}
is given by a diagonal matrix,
\begin{align}\label{eq:diagcovmat}
4 \boldsymbol{\Gamma}[\hat{\rho}_1\otimes\cdots\otimes\hat{\rho}_M,\hat{\mathbf{H}}]=4\begin{pmatrix}(\Delta\hat{H}_1)^2_{\hat{\rho}_1}& 0 & \cdots & 0\\
\vdots & &\ddots & \vdots\\
0 & \cdots & 0 &(\Delta\hat{H}_M)^2_{\hat{\rho}_M}\end{pmatrix},
\end{align}
which describes the covariance of the product state $\hat{\rho}_1\otimes\cdots\otimes\hat{\rho}_M$ of reduced density matrices. It can be obtained from the full covariance matrix $\boldsymbol{\Gamma}[\hat{\rho}_{\mathrm{m-sep}},\hat{\mathbf{H}}]$ by removing all the off-diagonal elements.

\subsection{State-independent bounds}
In the following we maximize the bound~(\ref{eq:diagcovmat}) over all quantum states under different conditions.

\subsubsection{Fluctuating number of particles in each mode}
We consider a generic state \cite{PezzePRA2015}
\begin{align}\label{eq:flucnumstate}
\hat{\rho}_k = \sum_{n=0}^N p_{k,n}|\Psi_{k,n}\rangle\langle\Psi_{k,n}|,
\end{align}
where $|\Psi_{k,n}\rangle$ is a state of $0\leq n \leq N$ particles in mode $k$, with $p_{k,n}  \geq 0$ and $\sum_{n=0}^N p_{k,n} = 1$.
The variance of a generic operator $\hat{H}_k$ is bounded by
\begin{align}\label{eq:upbndvarfn}
(\Delta\hat{H}_k)^2_{\hat{\rho}_k}&\leq \langle\hat{H}^2_k\rangle_{\hat{\rho}_k}\notag\\
&=\sum_{n=0}^N p_{k,n}\langle\Psi_{k,n}|\hat{H}^2_k|\Psi_{k,n}\rangle\notag\\
&\leq \lambda^2_{k\max} \sum_{n=0}^N p_{k,n} n^2 \notag \\
&=\lambda^2_{k\max}\langle\hat{N}_k^2\rangle_{\hat{\rho}_k},
\end{align}
where we used that $\hat{H}_k$ conserves the number of particles and that 
\begin{align}\label{eq:uprlimh2}
\langle\Psi_{k,n}|\hat{H}^2_k|\Psi_{k,n}\rangle\leq \Lambda_{k\max}^2 = n^2 \lambda_{k\max}^2, 
\end{align}
with $\Lambda_{k\max} = n \lambda_{k\max}$ the maximum eigenvalue of $\hat{H}_k$ and 
$\lambda_{k\max}$ the maximum eigenvalue of $\hat{h}_k^{(i)}$ (for all $i=1,...,n$).
The first inequality is saturated by states with $\langle\hat{H}_k\rangle_{\hat{\rho}_k}=0$ and the second by states with 
$\langle\Psi_{k,n}|\hat{H}^2_k|\Psi_{k,n}\rangle=n^2\lambda^2_{k\max}$ for all $\gamma$ and $n$. 
Both conditions can be satisfied for Hamiltonians with the property~(\ref{eq:property}) 
when the $|\Psi_{k,n}\rangle$ are NOON states with $n$ particles in a superposition of smallest and largest eigenvalue of $\hat{h}_k$.

By combining Eqs.~(\ref{eq:mseplimitmtx}) and~(\ref{eq:upbndvarfn}) we thus obtain the following upper sensitivity limit for mode-separable states, as a function of the fluctuations $\langle \hat{N}_k^2\rangle$:
\begin{align}\label{eq:MSbound}
\mathbf{F}_{\mathrm{MS}}= 4\begin{pmatrix}\lambda_{1\max}^2\langle\hat{N}_1^2\rangle& 0 & \cdots & 0\\
\vdots & &\ddots & \vdots\\
0 & \cdots & 0 &\lambda_{M\max}^2\langle\hat{N}_M^2\rangle\end{pmatrix}.
\end{align}

Finally, we notice that the state (\ref{eq:flucnumstate}) does not contain coherences between different numbers of particles. 
This is not a restriction: since the Hamiltonian $\hat{H}_k$ does not couple states with different numbers of particles, 
the same bound (\ref{eq:MSbound}) can be obtained by maximizing over single-mode state with number coherences \cite{PezzePRA2015}, 
$\hat{\rho}_{k}=\sum_{\gamma}p_{k,\gamma}\vert \Psi_{k,\gamma} \rangle\langle\Psi_{k,\gamma} \vert$ with $|\Psi_{k,\gamma}\rangle = \sum_{n=0}^N e^{i \phi_{k,\gamma,n}} \sqrt{Q_{k,\gamma,n}} |\Psi_{k,\gamma,n}\rangle$ and $n$-particle states $|\Psi_{k,\gamma,n}\rangle$ and arbitrary phases $\phi_{k,\gamma,n}$.

\subsubsection{Fixed number of particles in each mode}
For arbitrary Hamiltonians $\hat{H}_k$ whose spectrum has upper and lower bounds $\Lambda_{k+}$ and $\Lambda_{k-}$, respectively, we have
\begin{align}\label{eq:varvupbnd}
2(\Delta\hat{H}_k)_{\hat{\rho}}\leq\delta\Lambda_k\equiv\Lambda_{k+}-\Lambda_{k-}.
\end{align}
Using this in Eq.~(\ref{eq:diagcovmat}) yields $\mathbf{F}_{\mathrm{MS}}=\mathrm{diag}((\delta\Lambda_1)^2,\dots,(\delta\Lambda_M)^2)$ which is achieved by a mode product of superposition states $|\Psi_{\mathrm{MS}}\rangle=\bigotimes_{k=1}^M(|\Lambda_{k+}\rangle+|\Lambda_{k-}\rangle)/\sqrt{2}$.

If the number of particles in each mode is fixed and equal to $N_k$ the extremal eigenvalues of the $\hat{H}^{(N_k)}_k$ are given by $N_k\lambda_{k\pm}$ and we obtain
\begin{align}
\mathbf{F}_{\mathrm{MS}}=\begin{pmatrix}N_1^2(\lambda_{1+}-\lambda_{1-})^2& 0 & \cdots & 0\\
\vdots & &\ddots & \vdots\\
0 & \cdots & 0 &N_M^2(\lambda_{M+}-\lambda_{M-})^2\end{pmatrix}.\notag
\end{align}
This bound follows from Eq.~(\ref{eq:MSbound}) for $\langle\hat{N}^2_k\rangle=N_k^2$ when $(\lambda_{k+}-\lambda_{k-})^2 = 4\lambda_{k\max}^2$, i.e., for Hamiltonians with the condition~(\ref{eq:property}). It is saturated by a mode product of NOON states with $N_k$ particles for $k=1,\dots,M$, i.e.,
\begin{align}
|\Psi_{\mathrm{MsPe}}\rangle=\bigotimes_{k=1}^M \frac{ |N_k,\lambda_{k+}\rangle + |N_k,\lambda_{k-}\rangle}{\sqrt{2}},\notag
\end{align}
where $|N_k,\lambda_{k\pm}\rangle$ describes $N_k$ particles in the state with eigenvalue $\lambda_{k\pm}$. We recover a single-mode NOON state in the case $M=1$, which achieves the Heisenberg limit $F_Q[\frac{ |N_1,+\rangle + |N_1,-\rangle }{\sqrt{2}},\hat{H}_1]=N_1^2(\lambda_{1+}-\lambda_{1-})^2$ for single-parameter estimation \cite{GiovannettiPRL2006}.

\section{The weak multiparameter Cram\'er-Rao bound}
We demonstrate here the chain of inequalities
\begin{align}\label{eq:weakmCRLB}
\mathbf{n}^T\boldsymbol{\Sigma}\mathbf{n}\geq \frac{(\mathbf{n}^T\mathbf{n})^2}{\mathbf{n}^T\mathbf{F}\mathbf{n}}\geq \frac{(\mathbf{n}^T\mathbf{n})^2}{\mathbf{n}^T\mathbf{F}_Q\mathbf{n}},
\end{align}
valid for locally unbiased estimators and all $\mathbf{n}$. With the normalization condition $\mathbf{n}^T\mathbf{n}=1$, the bounds take on the form presented in the main manuscript. 

The first inequality identifies the weak multiparameter Cram\'er-Rao bound. Its proof assumes standard regularity conditions for $p(x|\boldsymbol{\theta})$ and its derivatives \cite{Kay1993}. The normalization condition $\int dx \, p(x|\boldsymbol{\theta})=1 $ implies $\int dx\frac{\partial \log p(x|\boldsymbol{\theta})}{\partial \theta_j}p(x|\boldsymbol{\theta})=0$
and
\begin{align}\label{eq:wcrb2}
\int dx \, \theta_i\frac{\partial \log p(x|\boldsymbol{\theta})}{\partial \theta_j}p(x|\boldsymbol{\theta})=0.
\end{align}
Furthermore, for locally unbiased estimators, we have $\int dx \, \theta_{\mathrm{est},i}(x) p(x|\boldsymbol{\theta}) = \theta_i$ and thus
\begin{align} \label{eq:wcrb1}
\int dx \, \theta_{\mathrm{est},i}(x) \frac{\partial \log p(x|\boldsymbol{\theta})}{\partial \theta_j} p(x|\boldsymbol{\theta}) = \delta_{ij}.
\end{align}
Taking the difference of Eqs.~(\ref{eq:wcrb1}) and~(\ref{eq:wcrb2}), we obtain
\begin{align}
\int dx \left(\theta_{\mathrm{est},i}(x)-\theta_i\right)\frac{\partial \log p(x|\boldsymbol{\theta})}{\partial \theta_j}p(x|\boldsymbol{\theta})=\delta_{ij}, \notag
\end{align}
or, in matrix form, 
\begin{align}
\int dx \left(\boldsymbol{\theta}_{\mathrm{est}}(x)-\boldsymbol{\theta}\right)\left(\frac{\partial \log p(x|\boldsymbol{\theta})}{\partial \boldsymbol{\theta}}\right)^Tp(x|\boldsymbol{\theta})=\mathbb{I}. \notag
\end{align}
For an arbitrary $\mathbf{n}\in\mathbb{R}^M$, we obtain
\begin{align}
\int dx \, \mathbf{n}^T\left(\boldsymbol{\theta}_{\mathrm{est}}(x)-\boldsymbol{\theta}\right)\left(\frac{\partial \log p(x|\boldsymbol{\theta})}{\partial \boldsymbol{\theta}}\right)^T\mathbf{n}p(x|\boldsymbol{\theta})=\mathbf{n}^T\mathbf{n}. \notag
\end{align}
The first inequality in~(\ref{eq:weakmCRLB}) is obtained from the Cauchy-Schwarz inequality:
\begin{align}
\left(\mathbf{n}^T\mathbf{n}\right)^2&\leq\left(\int dx\mathbf{n}^T\left(\boldsymbol{\theta}_{\mathrm{est}}(x)-\boldsymbol{\theta}\right)^2\mathbf{n}p(x|\boldsymbol{\theta})\right)\notag\\&\quad\times\left(\int dx\mathbf{n}^T\left(\frac{\partial \log p(x|\boldsymbol{\theta})}{\partial \boldsymbol{\theta}}\right)\left(\frac{\partial \log p(x|\boldsymbol{\theta})}{\partial \boldsymbol{\theta}}\right)^T\mathbf{n}p(x|\boldsymbol{\theta})\right)\notag\\
&=\left(\mathbf{n}^T\boldsymbol{\Sigma}\mathbf{n}\right)\left(\mathbf{n}^T\mathbf{F}\mathbf{n}\right).\notag
\end{align}
The weak quantum multiparameter Cram\'er-Rao bound then follows from $\mathbf{n}^T\mathbf{F}\mathbf{n}\leq \mathbf{n}^T\mathbf{F}_Q\mathbf{n}$, that can always be saturated by an optimal measurement, which, in general, may depend on $\mathbf{n}$. This follows since $\mathbf{n}^T\mathbf{F}_Q\mathbf{n}$ can be related to the maximal single-parameter sensitivity [Eq.~(\ref{eq:singleQFM})] which can always be achieved by an optimal measurement \cite{BraunsteinPRL1994}. In the scenario of this manuscript there is always an optimal measurement independent of $\mathbf{n}$ such that $\mathbf{n}^T\mathbf{F}\mathbf{n}=\mathbf{n}^T\mathbf{F}_Q\mathbf{n}$ holds.

Notice that the above derivation does not assume the existence of $\mathbf{F}^{-1}$. If $\mathbf{F}^{-1}$ exists, we further have the following ordering relation
\begin{align}\label{eq:chain}
\mathbf{n}^T\boldsymbol{\Sigma}\mathbf{n}\geq\mathbf{n}^T\mathbf{F}^{-1}\mathbf{n}\geq\frac{(\mathbf{n}^T\mathbf{n})^2}{\mathbf{n}^T\mathbf{F}\mathbf{n}}.
\end{align}
The first inequality is the Cram\'{e}r-Rao bound, the second follows immediately from the Cauchy-Schwarz inequality $(\mathbf{f}^T\mathbf{f})(\mathbf{g}^T\mathbf{g})\geq (\mathbf{f}^T\mathbf{g})^2$ with $\mathbf{f}=\sqrt{\mathbf{F}}\mathbf{n}$ and $\mathbf{g}=\sqrt{\mathbf{F}}^{-1}\mathbf{n}$ (note that $\sqrt{\mathbf{F}}$ exists since $\mathbf{F}>0$). The bound is saturated if and only if $\mathbf{n}$ is an eigenvector of $\mathbf{F}$. The same chain of inequalities~(\ref{eq:chain}) holds also for the quantum Fisher matrix: $\mathbf{n}^T\boldsymbol{\Sigma}\mathbf{n}\geq\mathbf{n}^T\mathbf{F}_Q^{-1}\mathbf{n}\geq(\mathbf{n}^T\mathbf{n})^2/\mathbf{n}^T\mathbf{F}_Q\mathbf{n}$. If the weak multiparameter quantum Cram\'er-Rao bound is saturated, i.e., if $\mathbf{n}^T\boldsymbol{\Sigma}\mathbf{n}=(\mathbf{n}^T\mathbf{n})^2/\mathbf{n}^T\mathbf{F}_Q\mathbf{n}$ then the stronger bound, if it exists, must coincide with the weaker bound, i.e., $\mathbf{n}^T\mathbf{F}_Q^{-1}\mathbf{n}=(\mathbf{n}^T\mathbf{n})^2/\mathbf{n}^T\mathbf{F}_Q\mathbf{n}$.

\section{Multiparameter Heisenberg limit}
\subsection{State-dependent bounds}
The Heisenberg limit is given by the maximal quantum Fisher matrix achievable by any state. The upper bound~(\ref{eq:QFMvar}), which is saturated by pure states, maps this problem to the maximization of the covariance matrix:
\begin{align}\label{eq:HLfirststep}
\mathbf{n}^T \mathbf{F}_Q[\hat{\rho},\hat{\mathbf{H}}] \mathbf{n}\leq 4\mathbf{n}^T \boldsymbol{\Gamma}[\hat{\rho},\hat{\mathbf{H}}] \mathbf{n}.
\end{align}
This state-dependent bound can be further improved using the following relation: 
\begin{align}\label{eq:covvarup}
|\mathrm{Cov}(\hat{H}_k,\hat{H}_l)_{\hat{\rho}}|\leq (\Delta \hat{H}_k)_{\hat{\rho}}(\Delta \hat{H}_l)_{\hat{\rho}}.
\end{align}
This bound expresses a necessary condition for the positive semi-definiteness of the covariance matrix. It can be derived using the Cauchy-Schwarz inequality for the Hilbert-Schmidt scalar product $\langle \hat{A}_1,\hat{A}_2\rangle=\mathrm{Tr}\{\hat{A}_1^{\dagger}\hat{A}_2\}$ and vectors $\hat{A}_i=(\hat{H}_i-\mathrm{Tr}\{\hat{H}_i\hat{\rho}\}\hat{\mathbb{I}})\sqrt{\hat{\rho}}$. We obtain $|\mathrm{Cov}(\hat{H}_k,\hat{H}_l)_{\hat{\rho}}|=|\mathrm{Re}\langle \hat{A}_1,\hat{A}_2\rangle|\leq |\langle \hat{A}_1,\hat{A}_2\rangle|\leq\sqrt{\langle \hat{A}_1,\hat{A}_1\rangle\langle \hat{A}_2,\hat{A}_2\rangle}=(\Delta \hat{H}_k)_{\hat{\rho}}(\Delta \hat{H}_l)_{\hat{\rho}}$. The first inequality is always saturated for commuting $\hat{H}_1$ and $\hat{H}_2$ as considered here. Equality in the second step is achieved if and only if there exists a constant $\alpha$, such that 
$(\hat{H}_1-\alpha\hat{H}_2)\sqrt{\hat{\rho}}=(\mathrm{Tr}\{\hat{H}_1\hat{\rho}\}-\alpha\mathrm{Tr}\{\hat{H}_2\hat{\rho}\})\sqrt{\hat{\rho}}$.

Using Eq.~(\ref{eq:covvarup}), we can further bound the right-hand side of Eq.~(\ref{eq:HLfirststep}) as
\begin{align}\label{eq:HLsecondstep}
\mathbf{n}^T\boldsymbol{\Gamma}[\hat{\rho},\hat{\mathbf{H}}]\mathbf{n}&=\sum_{k,l=1}^Mn_kn_l\mathrm{Cov}(\hat{H}_k,\hat{H}_l)_{\hat{\rho}}\notag\\
&\leq \sum_{k,l=1}^M|n_kn_l|\left|\mathrm{Cov}(\hat{H}_k,\hat{H}_l)_{\hat{\rho}}\right|\notag\\
&\leq \sum_{k,l=1}^M|n_kn_l|(\Delta \hat{H}_k)_{\hat{\rho}}(\Delta \hat{H}_l)_{\hat{\rho}}\notag\\
&\leq \sum_{k,l=1}^Mn_kn_l\mathrm{sgn}(n_k)\mathrm{sgn}(n_l)(\Delta \hat{H}_k)_{\hat{\rho}}(\Delta \hat{H}_l)_{\hat{\rho}}. 
\end{align}
Together with Eq.~(\ref{eq:HLfirststep}) this leads to the state-dependent sensitivity bound for arbitrary $\mathbf{n}$:
\begin{align}\label{eq:statedepHL}
\mathbf{n}^T \mathbf{F}_Q[\hat{\rho},\hat{\mathbf{H}}] \mathbf{n} \leq 4\mathbf{n}^T \boldsymbol{\Gamma}^{\mathbf{n}}[\hat{\rho},\hat{\mathbf{H}}]
\mathbf{n}.
\end{align}
The bound can be written as $\boldsymbol{\Gamma}^{\mathbf{n}}[\hat{\rho},\hat{\mathbf{H}}]
= \mathbf{v}_{\hat{\rho}}^{\mathbf{n}} \mathbf{v}_{\hat{\rho}}^{\mathbf{n}T}$, where $\mathbf{v}_{\hat{\rho}}^{\mathbf{n}}$ is a vector with elements $\epsilon_k(\Delta \hat{H}_k)_{\hat{\rho}}$, for $k=1,\dots,M$ and $\epsilon_k=\mathrm{sgn}(n_k)$. 

\subsection{State-independent bounds}
In the following we maximize the bound~(\ref{eq:statedepHL}) over arbitrary quantum states under different conditions.

\subsubsection{Fluctuating number of particles in each mode}
We write a generic quantum state of $N$ particles in $M$ modes as $\hat{\rho}=\sum_{\gamma}p_{\gamma}|\Psi_{\gamma}\rangle\langle\Psi_{\gamma}|$, with $|\Psi_{\gamma}\rangle=\sum_{\mathbf{N}}\sqrt{Q_{\gamma,\mathbf{N}}}|\Psi_{\gamma,\mathbf{N}}\rangle$, where $\mathbf{N}=(N_1,\dots,N_M)$ is a vector of fixed particle numbers for all modes and the sum extends over all possible combinations with $\sum_{k=1}^MN_k=N$. We allow for particle coherence among the different modes. The covariance is bounded as
\begin{align}\label{eq:upbndcovfn}
|\mathrm{Cov}(\hat{H}_k,\hat{H}_l)_{\hat{\rho}}|&\leq |\langle\hat{H}_k\hat{H}_l\rangle_{\hat{\rho}}|\notag\\
&\leq\sum_{\gamma}p_{\gamma}\sum_{\mathbf{N}}|Q_{\gamma,\mathbf{N}}||\langle\Psi_{\gamma,\mathbf{N}}|\hat{H}_k\hat{H}_l|\Psi_{\gamma,\mathbf{N}}\rangle|\notag\\
&\leq \sum_{\gamma}p_{\gamma}\sum_{N_k=0}^{N}|Q_{\gamma,\mathbf{N}}|N_kN_l\lambda_{k\max}\lambda_{l\max}\notag\\
&=\lambda_{k\max}\lambda_{l\max}\sum_{\gamma}p_{\gamma}\langle\hat{N}_k\hat{N}_l\rangle_{|\Psi_{\gamma}\rangle}\notag\\
&=\lambda_{k\max}\lambda_{l\max}\langle\hat{N}_k\hat{N}_l\rangle_{\hat{\rho}}.
\end{align}
The first inequality is saturated if and only if $\langle\hat{H}_k\rangle_{\hat{\rho}}=0$ for all $k$. In the second we used the triangle inequality and again that the $\hat{H}_k$ conserve the number of particles. The third inequality follows from the Cauchy-Schwarz inequality yielding $|\langle\Psi_{\gamma,\mathbf{N}}|\hat{H}_k\hat{H}_l|\Psi_{\gamma,\mathbf{N}}\rangle|\leq\sqrt{\langle\Psi_{\gamma,\mathbf{N}}|\hat{H}_k^2|\Psi_{\gamma,\mathbf{N}}\rangle\langle\Psi_{\gamma,\mathbf{N}}|\hat{H}_l^2|\Psi_{\gamma,\mathbf{N}}\rangle}$ and then using Eq.~(\ref{eq:uprlimh2}). Combining this with Eq.~(\ref{eq:HLfirststep}) and $\mathbf{n}^T\boldsymbol{\Gamma}[\hat{\rho},\hat{\mathbf{H}}]\mathbf{n}\leq \sum_{k,l=1}^M|n_kn_l|\left|\mathrm{Cov}(\hat{H}_k,\hat{H}_l)_{\hat{\rho}}\right|$ [see Eq.~(\ref{eq:HLsecondstep})] leads the bound $\mathbf{n}^T\mathbf{F}_Q[\hat{\rho},\hat{\mathbf{H}}]\mathbf{n}\leq \mathbf{n}^T\mathbf{F}_{\mathrm{HL'}}^{\mathbf{n}}\mathbf{n}$, 
where
\begin{align}
\mathbf{F}_{\mathrm{HL'}}^{\mathbf{n}}&= 4\begin{pmatrix} \lambda_{1\max}^2\langle\hat{N}_1^2\rangle & \dots & \tilde{\lambda}_{1\max}\tilde{\lambda}_{M\max}\langle\hat{N}_1\hat{N}_M\rangle\\
\vdots & \ddots & \vdots\\
\tilde{\lambda}_{1\max}\tilde{\lambda}_{M\max}\langle\hat{N}_1\hat{N}_M\rangle &  \cdots  & \lambda_{M\max}^2\langle\hat{N}_M^2\rangle
\end{pmatrix},\notag
\end{align}
and $\tilde{\lambda}_{k\max} = \mathrm{sgn}(n_k)\lambda_{k\max}$. The upper bound is attained by pure states with $\langle\hat{H}_k\rangle_{|\Psi\rangle}=0$ and $\langle\hat{H}_k\hat{H}_l\rangle_{|\Psi\rangle} = \tilde{\lambda}_{k\max}\tilde{\lambda}_{l\max}\langle \hat{N}_k\hat{N}_l\rangle$, for all $k,l=1,\dots,M$, which can be achieved for interferometers with the property~(\ref{eq:property}) by optimal states discussed below in Eq.~(\ref{eq:psin}).

Using the further Cauchy-Schwarz inequality $\langle\hat{N}_k\hat{N}_l\rangle_{\hat{\rho}}\leq\sqrt{\langle\hat{N}_k^2\rangle_{\hat{\rho}}\langle\hat{N}_l^2\rangle_{\hat{\rho}}}$ in Eq.~(\ref{eq:upbndcovfn}), we obtain $\mathbf{n}^T\mathbf{F}_Q[\hat{\rho},\hat{\mathbf{H}}]\mathbf{n}\leq \mathbf{n}^T\mathbf{F}_{\mathrm{HL}}^{\mathbf{n}}\mathbf{n}$ with $\mathbf{F}_{\mathrm{HL}}^{\mathbf{n}}=\mathbf{v}^{\mathbf{n}}\mathbf{v}^{\mathbf{n}T}$. The vector $\mathbf{v}^{\mathbf{n}}=2(\tilde{\lambda}_{1\max}\sqrt{\langle\hat{N}_1^2\rangle},\dots,\tilde{\lambda}_{M\max}\sqrt{\langle\hat{N}_M^2\rangle})$ is determined by the single-mode expectation values $\langle\hat{N}_k^2\rangle$. This bound was given in the main manuscript for the case $\lambda_{k\max}=\frac{1}{2}$ for all $k$.

\subsubsection{Fixed number of particles in each mode}
When the spectrum of the $\hat{H}_k$ is bounded, state-independent upper sensitivity limits can be obtained by using Eq.~(\ref{eq:varvupbnd}) in~(\ref{eq:HLsecondstep}). We obtain $\mathbf{n}^T\mathbf{F}_Q[\hat{\rho}]\mathbf{n}\leq \mathbf{n}^T\mathbf{F}_{\mathrm{HL,b}}^{\mathbf{n}}\mathbf{n}$ where
\begin{align}\label{eq:HLbnd}
\mathbf{F}_{\mathrm{HL,b}}^{\mathbf{n}}=\mathbf{f}^{\mathbf{n}}\mathbf{f}^{\mathbf{n}T}
\end{align}
with $\mathbf{f}^{\mathbf{n}}=(\epsilon_1\delta\Lambda_1,\dots,\epsilon_M\delta \Lambda_M)$. Using the decomposition~(\ref{eq:Hksumi}) into single-particle Hamiltonians, we obtain $\delta \Lambda_k=N_k(\lambda_{k+}-\lambda_{k-})$, where $N_k$ is the number of particles in mode $k$. This bound coincides with $\mathbf{F}_{\mathrm{HL}}^{\mathbf{n}}$ for $\langle\hat{N}_k^2\rangle=N_k^2$ when $4\lambda_{k\max}^2=(\lambda_{k+}-\lambda_{k-})^2$, i.e., for Hamiltonians with the property~(\ref{eq:property}).

\subsubsection{Optimal states}
We now show that for any $\mathbf{n}$, there exists a family of quantum states whose quantum Fisher matrix coincides with $\mathbf{F}_{\mathrm{HL}}^{\mathbf{n}}$. Let us denote by $|N_k,\lambda_{k\pm\epsilon_k}\rangle$ a quantum state with $N_k$ particles in the eigenstate $|\lambda_{k\pm\epsilon_k}\rangle$ of mode $k$, where $\epsilon_k=\mathrm{sgn}(n_k)$. For example, if $n_k$ is positive, $|\lambda_{k\pm\epsilon_k}\rangle$ yields $|\lambda_{k+}\rangle$, whereas if $n_k$ is negative we obtain $|\lambda_{k-}\rangle$, where $\lambda_{\pm}$ are the largest and smallest eigenvalue of $\hat{h}^{(i)}_k$ as before [recall Eq.~(\ref{eq:Hksumi})]. Now, consider the family of states
\begin{align}\label{eq:psin}
|\Psi^{\mathbf{n}}_{\mathrm{MePe}}\rangle&=\frac{1}{\sqrt{2}}(|N_1,\lambda_{1+\epsilon_1}\rangle\otimes|N_2,\lambda_{2+\epsilon_2}\rangle\otimes\dots\otimes|N_M,\lambda_{M+\epsilon_M}\rangle\notag\\&\quad+|N_1,\lambda_{1-\epsilon_1}\rangle\otimes|N_2,\lambda_{2-\epsilon_2}\rangle\otimes\dots\otimes|N_M,\lambda_{M-\epsilon_M}\rangle).
\end{align}
These states have the property $\langle\Psi^{\mathbf{n}}_{\mathrm{MePe}}|\hat{H}_k|\Psi^{\mathbf{n}}_{\mathrm{MePe}}\rangle=\frac{1}{2}N_k(\lambda_{k+\epsilon_k}+\lambda_{k-\epsilon_k})$, and $\langle\Psi^{\mathbf{n}}_{\mathrm{MePe}}|\hat{H}_k\hat{H}_l|\Psi^{\mathbf{n}}_{\mathrm{MePe}}\rangle=\frac{1}{2}N_kN_l(\lambda_{+\epsilon_k}\lambda_{+\epsilon_l}+\lambda_{-\epsilon_k}\lambda_{-\epsilon_l})$. This leads to
\begin{align}
\mathrm{Cov}(\hat{H}_k,\hat{H}_l)_{|\Psi^{\mathbf{n}}_{\mathrm{MePe}}\rangle}&=\frac{1}{2}N_kN_l(\lambda_{k+\epsilon_k}\lambda_{l+\epsilon_l}+\lambda_{k-\epsilon_k}\lambda_{l-\epsilon_l})\notag\\&\quad-\frac{1}{4}N_kN_l(\lambda_{k+\epsilon_k}+\lambda_{k-\epsilon_k})(\lambda_{l+\epsilon_l}+\lambda_{l-\epsilon_l})\notag\\
&=\frac{1}{4}N_kN_l(\lambda_{k+\epsilon_k}-\lambda_{k-\epsilon_k})(\lambda_{l+\epsilon_l}-\lambda_{l-\epsilon_l})\notag\\
&=\frac{1}{4}\epsilon_k\epsilon_l(\delta\Lambda_{k})(\delta\Lambda_{l}).\notag
\end{align}
Hence, these states saturate the bounds~(\ref{eq:statedepHL}) and~(\ref{eq:HLbnd}) and hence, if~(\ref{eq:property}) holds, their quantum Fisher matrix coincides with $\mathbf{F}_{\mathrm{HL'}}^{\mathbf{n}}$ and $\mathbf{F}_{\mathrm{HL}}^{\mathbf{n}}$.

\subsection{Stepwise enhancement through particle and mode entanglement}
\subsubsection{Entanglement among a subset of modes}
The derivation of Eq.~(\ref{eq:mseplimitmtx}) can be extended to states that are separable in a specific partition $\Lambda=\mathcal{A}_1|\dots|\mathcal{A}_L$, where the $\mathcal{A}_m$ describe a coarse-grained ensemble of modes, i.e., states that allow for a decomposition of the type \cite{Quantum2017}
\begin{align}
\hat{\rho}_{\Lambda-\mathrm{sep}}=\sum_{\gamma}p_{\gamma}\hat{\rho}_{\gamma,\mathcal{A}_1}\otimes\dots\otimes\hat{\rho}_{\gamma,\mathcal{A}_L},\notag
\end{align}
where the $\hat{\rho}_{\gamma,\mathcal{A}_m}$ are density matrices on $\mathcal{A}_m$. This yields the block-diagonal upper sensitivity limit
\begin{align}
\mathbf{F}_Q[\hat{\rho}_{\Lambda-\mathrm{sep}},\hat{\mathbf{H}}]\leq 4 \boldsymbol{\Gamma}[\hat{\rho}_{\mathcal{A}_1}\otimes\cdots\otimes\hat{\rho}_{\mathcal{A}_L},\hat{\mathbf{H}}],\notag
\end{align}
with the reduced density matrices $\hat{\rho}_{\mathcal{A}_m}=\sum_{\gamma}p_{\gamma}\hat{\rho}_{\gamma,\mathcal{A}_m}$ for $\mathcal{A}_m$. By maximizing the block-diagonal covariance matrix in an analog way as before, we obtain the sensitivity limit for $\Lambda$-separable states. It can be obtained from the Heisenberg limit by removing those off-diagonal blocks that describe correlations between different $\mathcal{A}_m$.

\subsubsection{Entanglement among a subset of particles in each mode}
Let us now consider the case of a fixed and integer number of particles $N_k$ in each mode, of which not more than $1\leq P_k\leq N_k$ are entangled. The amount of particle entanglement in all modes is characterized by the vector $\mathbf{P}=(P_1,\dots,P_M)$ and we call states $\hat{\rho}_{\mathbf{P}-\mathrm{prod}}$ with limited particle entanglement $\mathbf{P}$-producible. We allow for entanglement among all modes. The sensitivity bounds can be derived directly from the quantum Fisher matrix using similar steps as those that led to the Heisenberg limit. From the Cauchy-Schwarz inequality we have $|(\mathbf{F}_Q)_{kl}|\leq\sqrt{(\mathbf{F}_Q)_{kk}(\mathbf{F}_Q)_{ll}}$ (an analogous relation holds for the elements of $\mathbf{F}$) for the elements of $\mathbf{F}_Q$. In analogy to Eq.~(\ref{eq:HLsecondstep}), this yields the state-dependent bound $\mathbf{n}^T\mathbf{F}_Q[\hat{\rho}_{\mathbf{P}-\mathrm{prod}},\hat{\mathbf{H}}]\mathbf{n}
\leq \mathbf{n}^T\mathbf{F}_{\hat{\rho}_{\mathbf{P}-\mathrm{prod}}}^{\mathbf{n}}\mathbf{n}$,
with $\mathbf{F}_{\hat{\rho}_{\mathbf{P}-\mathrm{prod}}}^{\mathbf{n}}=\mathbf{v}_{\hat{\rho}_{\mathbf{P}-\mathrm{prod}}}^{\mathbf{n}}\mathbf{v}_{\hat{\rho}_{\mathbf{P}-\mathrm{prod}}}^{\mathbf{n}T}$ and $\mathbf{v}_{\hat{\rho}_{\mathbf{P}-\mathrm{prod}}}^{\mathbf{n}}$ is a vector with elements $\epsilon_k\sqrt{F_Q[\hat{\rho}_{\mathbf{P}-\mathrm{prod}},\hat{H}_k]}$ for $k=1,\dots,M$. The single-parameter sensitivity for $N$-particle states $\hat{\rho}_{P_k-\mathrm{prod}}$ with no more than $P_k$ entangled particles is bounded by \cite{HyllusPRA2012} $F_Q[\hat{\rho}_{P_k-\mathrm{prod}},\hat{H}_k]\leq (s_kP_k^2+r_k^2)(\lambda_{k+}-\lambda_{k-})^2$, where $s_k=\lfloor N_k/P_k\rfloor$ and $r_k=N_k-s_kP_k$. Note that $N_k\leq(s_kP_k^2+r_k^2)\leq N_k^2$. This yields the state-independent bound
\begin{align}\label{eq:piprod}
\mathbf{n}^T\mathbf{F}_Q[\hat{\rho}_{\mathbf{P}-\mathrm{prod}},\hat{\mathbf{H}}]\mathbf{n}
\leq \mathbf{n}^T\mathbf{F}_{\mathbf{P}}^{\mathbf{n}}\mathbf{n},
\end{align}
with $\mathbf{F}_{\mathbf{P}}^{\mathbf{n}}=\mathbf{v}_{\mathbf{P}}^{\mathbf{n}}\mathbf{v}_{\mathbf{P}}^{\mathbf{n}T}$ and $\mathbf{v}_{\mathbf{P}}^{\mathbf{n}}$ is a vector of elements $\sqrt{(s_kP_k^2+r_k^2)}(\lambda_{k+\epsilon_k}-\lambda_{k-\epsilon_k})$ for $k=1,\dots,M$.

\subsubsection{Multi-particle and multi-mode entanglement}
The results on particle and mode entanglement can be combined. Separability between specific modes leads to zero entries in the off-diagonal blocks that describe these mode correlations in the quantum Fisher matrix. We can therefore obtain the sensitivity bounds for states that are both $\mathbf{P}$-producible and $\Lambda$-separable by removing the off-diagonal blocks from~(\ref{eq:piprod}) that describe correlations across different groups of modes contained in $\Lambda$.

Let us quantify the quantum gain in Eq.~(\ref{eq:weakmCRLB}) for $N_k = \bar{N}= N/M$ particles in each mode, where $\bar{N}$ is assumed integer. We indicate as $\hat{\rho}_{M_e,P_e}$ states with 
not more than $P_e\leq N/M$ entangled particles in each mode and not more than $M_e\leq M$ entangled modes.  The achievable sensitivity for such a quantum state $\hat{\rho}_{M_e,P_e}$ is given by
\begin{align}
S_{M_e,P_e}=\mathbf{n}^T\mathbf{F}_Q[\hat{\rho}_{M_e,P_e},\hat{\mathbf{H}}]\mathbf{n}=\sum_{\mathcal{A}_m\in\Lambda}\sum_{kl\in\mathcal{A}_m}|n_kn_l|(sP_e^2+r^2),\notag
\end{align}
where $s=\lfloor N/(P_e M)\rfloor$, $r=N/M-sP_e$, and we assume $|\mathbf{n}|^2=1$. From the Cauchy-Schwarz inequality and using $|n_k|\leq 1$, we obtain $(\sum_{k=1}^K|n_k|^2)^2\leq\sum_{kl=1}^K|n_kn_l|\leq (\sum_{k=1}^K|n_k|^2)K$, where the lower bound is reached for $n_k=\delta_{kk_0}$ for some $k_0$ and the upper bound is achieved when $|n_k|=const$. Choosing $|n_k|=1/\sqrt{M}$ leads to $\sum_{\mathcal{A}_m\in\Lambda}\sum_{kl\in\mathcal{A}_m}|n_kn_l|= \sum_{\mathcal{A}_m\in\Lambda}M_m^2/M$, where $M_m$ is the number of modes in $\mathcal{A}_m$ and $\sum_{\mathcal{A}_m\in\Lambda}M_m=M$. This quantity is maximized by employing $u=\lfloor M/M_e\rfloor$ sets of $M_e$ entangled modes and a single set of the remaining $v=M-uM_e$ entangled modes, yields $\sum_{\mathcal{A}_m\in\Lambda}M_m^2/M\leq (uM_e^2+v^2)/M$ and
\begin{align}
S_{M_e,P_e}\leq S^{\max}_{M_e,P_e}\equiv (sP_e^2+r^2)(uM_e^2+v^2)/M.\notag
\end{align}

The absence of particle entanglement in each mode implies $P=1$ and $s=N/M$ (notice that this does not imply that all particles are separable as two particles that enter different modes may be entangled unless also mode entanglement is excluded). Full multiparticle entanglement in each mode is described by the case $P_e=N/M$, yielding $u=1$. Full mode separability means $M_e=1$ and $s=M$ and for full multimode entanglement we have $M_e=M$ and $u=1$. In all these cases $r=v=0$. This leads to the maximal sensitivities:
\begin{eqnarray}
S^{\max}_{1,1}&=N, \quad S^{\max}_{1,\frac{N}{M}}&=\frac{N^2}{M},\notag\\
S^{\max}_{M,1}&=NM, \quad S^{\max}_{M,\frac{N}{M}}&=N^2.\notag
\end{eqnarray}
The gain factor which was introduced in the main manuscript is obtained by normalizing the sensitivities with respect to the $S^{\max}_{1,1}$ level.

\subsection{Beyond a finite number of particles}
Entanglement between particles can only be defined for quantum states with a fixed, finite number of particles or mixtures thereof. However, mode entanglement can also exist when the total number of particles is not fixed, as frequently encountered in continuous-variable systems. The bounds for arbitrary mode-separable and mode-entangled states can be derived analogously when the condition of a fixed number of particles is relaxed and they coincide with those presented in this manuscript. Our results on mode entanglement therefore also apply to the case of bosonic particles and continuous-variable systems, described by local Hamiltonians of the form $\hat{H}_k=\sum_j\lambda_{kj}\hat{a}^{\dagger}_{kj}\hat{a}_{kj}$, where $\hat{a}_{kj}$ is a bosonic annihilation operator.

\section{Mode transformations and generic weight matrices}
For general states we have derived bounds on $\mathbf{n}^T\boldsymbol{\Sigma}\mathbf{n}$ for arbitrary $\mathbf{n}$. This corresponds to the figure of merit $\mathrm{Tr}\{\mathbf{W}\boldsymbol{\Sigma}\}$ with a rank-$1$ weight matrix $\mathbf{W}=\mathbf{n}\mathbf{n}^T$. Our results on mode-separable states are given in terms of matrix inequalities and thus imply bounds for arbitrary positive semidefinite $\mathbf{W}$. In other words, taking the spectral decomposition $\mathbf{W}=\sum_k w_k \mathbf{n}_k \mathbf{n}^T_k$ with $w_k \geq 0$, we have 
$\mathrm{Tr}\{\mathbf{W}\boldsymbol{\Sigma}\} = \sum_k w_k \mathbf{n}_k^T \boldsymbol{\Sigma} \mathbf{n}_k$ and 
$\mathbf{n}_k^T \boldsymbol{\Sigma} \mathbf{n}_k \geq \mathbf{n}_k^T \mathbf{F}_{\rm MS}^{-1} \mathbf{n}_k$ holds for all $\mathbf{n}_k$ and is saturated by an optimal state that does not depend on $\mathbf{n}_k$.
Thus, $\mathrm{Tr}\{\mathbf{W}\boldsymbol{\Sigma}\} \geq \mathrm{Tr}\{\mathbf{W}\mathbf{F}_{\rm MS}^{-1}\}$ is a saturable bound that holds for all mode-separable states. Yet, this is not the case for the Heisenberg limit, since the bound 
$\mathbf{n}^T_k \boldsymbol{\Sigma} \mathbf{n}_k \geq (\mathbf{n}^T_k \mathbf{F}_{\rm HL}^{\mathbf{n}_k} \mathbf{n}_k)^{-1}$ and the corresponding optimal state depend on $\mathbf{n}_k$. Here we show how all our results, including the Heisenberg limit, can be generalized to arbitrary weight matrices $\mathbf{W}$.

\subsection{Diagonal weight matrix}
Let us first consider the case of a diagonal weight matrix $\mathbf{W}=\sum_{k=1}^Mw_k\mathbf{e}_k\mathbf{e}_k^T$, where the $\mathbf{e}_k$ are the elements of the canonical basis. With Eq.~(\ref{eq:weakmCRLB}), we obtain the sensitivity bound
\begin{align}\label{eq:trS}
\mathrm{Tr}\{\mathbf{W}\boldsymbol{\Sigma}\}=\sum_{k=1}^Mw_k(\boldsymbol{\Sigma})_{kk}\geq\sum_{k=1}^Mw_k\frac{1}{\mathbf{F}_Q[\hat{\rho},\hat{\mathbf{H}}]_{kk}},
\end{align}
which, according to Eq.~(\ref{eq:chain}), is saturated if and only if $\mathbf{F}_Q[\hat{\rho},\hat{\mathbf{H}}]$ is diagonal in the canonical basis. Hence, for to this figure of merit, which contains no parameter-correlations, the highest sensitivity is achieved by a mode-separable state. We have $\mathrm{Tr}\{\mathbf{W}\boldsymbol{\Sigma}\}\geq\mathrm{Tr}\{\mathbf{W}\boldsymbol{\Sigma}^{\mathbf{W}}_{\max}\}$, where
\begin{align}\label{eq:diagGlimit}
\boldsymbol{\Sigma}^{\mathbf{W}}_{\max}\equiv\mathbf{F}^{-1}_{\mathrm{diag}}[\hat{\mathbf{H}}]
\end{align}
for diagonal $\mathbf{W}$ with $\mathbf{F}_{\mathrm{diag}}[\hat{\mathbf{H}}]=\max_{|\Psi_1\rangle\otimes\cdots\otimes|\Psi_M\rangle}\mathbf{F}_Q[|\Psi_1\rangle\otimes\cdots\otimes|\Psi_M\rangle,\hat{\mathbf{H}}]$ and we made use of the convexity of $\mathbf{F}_Q$. If $\mathbf{F}_{\mathrm{diag}}[\hat{\mathbf{H}}]$ is not invertible, the bound~(\ref{eq:trS}) may still be optimized by individually maximizing those $\mathbf{F}_Q[\hat{\rho},\hat{\mathbf{H}}]_{kk}=F_Q[\hat{\rho},\hat{H_k}]$ with a non-zero $w_k$.

We can extend this result to arbitrary matrices $\mathbf{W}$ by employing a mode transformation which diagonalizes $\mathbf{W}$. To see this, we first need to understand the transformation properties of the Fisher and covariance matrices.

\subsection{Transformation of parameters and generators}
The unitary evolution~(\ref{eq:modeevolution}) is determined by the scalar product $\hat{\mathbf{H}}\cdot\boldsymbol{\theta}$ which is invariant under the orthogonal transformations, i.e., $\hat{\mathbf{H}}\cdot\boldsymbol{\theta}=\hat{\mathbf{H}}^T\mathbf{O}^T\mathbf{O}\boldsymbol{\theta}=\hat{\mathbf{H}}'\cdot\boldsymbol{\theta}'$, where $\hat{\mathbf{H}}'=\mathbf{O}\hat{\mathbf{H}}$ and $\boldsymbol{\theta}'=\mathbf{O}\boldsymbol{\theta}$ are transformed vectors of generators and parameters, respectively.

How can the multiparameter sensitivity of an estimation of $\boldsymbol{\theta}'$ be related to that of $\boldsymbol{\theta}$? The answer is provided by the bilinearity of the covariance matrix~(\ref{eq:bilcov}), which yields $\boldsymbol{\Sigma}'=\mathbf{O}\boldsymbol{\Sigma}\mathbf{O}^T$. The next question is, how can the quantum Cram\'er-Rao bound for $\boldsymbol{\Sigma}'$, i.e., the quantum Fisher matrix for generators $\hat{\mathbf{H}}'$ be related to that of $\hat{\mathbf{H}}$? Using the spectral decomposition of $\hat{\rho}$, we obtain
\begin{align}\label{eq:transformQFI}
&\quad\left(\mathbf{F}_Q[\hat{\rho},\hat{\mathbf{H}}']\right)_{ij}\notag\\&=2\sum_{k,k'}\frac{(p_k-p_{k'})^2}{p_k+p_{k'}}\langle k|\hat{H}'_i|k'\rangle\langle k'|\hat{H}'_j|k\rangle\notag\\
&=2\sum_{k,k'}\frac{(p_k-p_{k'})^2}{p_k+p_{k'}}\langle k|\left(\sum_{l=1}^Mo_{il}\hat{H}_l\right)|k'\rangle\langle k'|\left(\sum_{l'=1}^Mo_{jl'}\hat{H}_{l'}\right)|k\rangle\notag\\
&=2\sum_{l,l'=1}^Mo_{il}o_{jl'}\sum_{k,k'}\frac{(p_k-p_{k'})^2}{p_k+p_{k'}}\langle k|\hat{H}_l|k'\rangle\langle k'|\hat{H}_{l'}|k\rangle\notag\\
&=\left(\mathbf{O}\mathbf{F}_Q[\hat{\rho},\hat{\mathbf{H}}]\mathbf{O}^T\right)_{ij},
\end{align}
which generalizes Eq.~(\ref{eq:singleQFM}) and we denoted the matrix elements of $\mathbf{O}$ as $o_{kl}$. Notice that this property holds for arbitrary matrices $\mathbf{O}$. The transformed multiparameter QCRB
\begin{align}\label{eq:trbound}
\boldsymbol{\Sigma}'\geq (\mathbf{F}_Q[\hat{\rho},\hat{\mathbf{H}}'])^{-1},
\end{align}
can directly be obtained by multiplying the bound $\boldsymbol{\Sigma}\geq (\mathbf{F}_Q[\hat{\rho},\hat{\mathbf{H}}])^{-1}$ with $\mathbf{O}$ and $\mathbf{O}^T$ from left and right and using $\mathbf{O}(\mathbf{F}_Q[\hat{\rho},\hat{\mathbf{H}}])^{-1}\mathbf{O}^T = (\mathbf{O}\mathbf{F}_Q[\hat{\rho},\hat{\mathbf{H}}]\mathbf{O}^T)^{-1} = (\mathbf{F}_Q[\hat{\rho},\hat{\mathbf{H}}'])^{-1}$. The last equality follows from the the transformation property~(\ref{eq:transformQFI}). The first equality follows since $(\mathbf{O}\mathbf{F}\mathbf{O}^T)^{-1}=\mathbf{O}\mathbf{F}^{-1}\mathbf{O}^T$, is true for arbitrary orthogonal matrices $\mathbf{O}$ and invertible matrices $\mathbf{F}$. To prove this, let $\mathbf{X}=(\mathbf{O}\mathbf{F}\mathbf{O}^T)^{-1}$ denote the inverse matrix of $\mathbf{O}\mathbf{F}\mathbf{O}^T$. Then, $\mathbf{X}\mathbf{O}\mathbf{F}\mathbf{O}^T=\mathbf{I}$ implies $\mathbf{O}^T\mathbf{X}\mathbf{O}\mathbf{F}=\mathbf{I}$ and $\mathbf{O}^T\mathbf{X}\mathbf{O}=\mathbf{F}^{-1}$. Finally we have $\mathbf{X}=\mathbf{O}\mathbf{F}^{-1}\mathbf{O}^T$, which proves the statement.

The sensitivity bounds can be transformed analogously and we find that whenever $\mathbf{F}_Q[\hat{\rho},\hat{\mathbf{H}}]\leq \mathbf{F}_{\max}$ holds, we have $\mathbf{F}_Q[\hat{\rho},\hat{\mathbf{H}}']\leq \mathbf{O} \mathbf{F}_{\max}\mathbf{O}^T$.

\subsection{Heisenberg limit for generic weight matrices}
To extend our result~(\ref{eq:diagGlimit}) to arbitrary positive semi-definite weight matrices $\mathbf{G}$, we use the spectral decomposition $\mathbf{W}=\mathbf{O}\mathbf{D}\mathbf{O}^T$, where $O$ is an orthogonal matrix and $\mathbf{W}=\mathrm{diag}(w_1,\dots,w_M)$ with $w_k\geq 0$. We obtain
\begin{align}\label{eq:genericG}
\mathrm{Tr}\{\mathbf{W}\boldsymbol{\Sigma}\}=\mathrm{Tr}\{\mathbf{D}\mathbf{O}^T\boldsymbol{\Sigma}\mathbf{O}\}=\sum_{k=1}^Mw_k(\boldsymbol{\Sigma}')_{kk},
\end{align}
where $\boldsymbol{\Sigma}'=\mathbf{O}^T\boldsymbol{\Sigma}\mathbf{O}$ describes the transformed covariance matrix. This transformed covariance matrix describes the sensitivity for a redefined set of parameters and is bounded by~(\ref{eq:trbound}) in terms of the quantum Fisher matrix for a transformed set of generators. Since $\mathbf{D}$ is diagonal, we can use Eq.~(\ref{eq:diagGlimit}) to obtain $\mathrm{Tr}\{\mathbf{D}\boldsymbol{\Sigma}'\}\geq \mathrm{Tr}\{\mathbf{D}\boldsymbol{\Sigma}'^{\mathbf{D}}_{\max}\}$ where $\boldsymbol{\Sigma}'^{\mathbf{D}}_{\max}=\mathbf{F}_{\mathrm{diag}}[\hat{\mathbf{H}}']^{-1}$ and $\mathbf{F}_{\mathrm{diag}}[\hat{\mathbf{H}}']$ is a quantum Fisher matrix that is diagonal in the modes $\hat{\mathbf{H}}'$ and maximized under the given constraints. Together with $\mathrm{Tr}\{\mathbf{W}\boldsymbol{\Sigma}\}=\mathrm{Tr}\{\mathbf{D}\boldsymbol{\Sigma}'\}$ this defines the sensitivity limit as
$\mathrm{Tr}\{\mathbf{W}\boldsymbol{\Sigma}\}\geq\mathrm{Tr}\{\mathbf{W}\boldsymbol{\Sigma}^{\mathbf{W}}_{\max}\}$ with
\begin{align}
\boldsymbol{\Sigma}^{\mathbf{W}}_{\max}\equiv \mathbf{O}\mathbf{F}_{\mathrm{diag}}[\hat{\mathbf{H}}']^{-1}\mathbf{O}^T,\notag
\end{align}
for arbitrary $\mathbf{W}\geq 0$, diagonalized by $\mathbf{O}$. Notice also that $\mathbf{O}\mathbf{F}_{\mathrm{diag}}[\hat{\mathbf{H}}']^{-1}\mathbf{O}^T=(\mathbf{O}\mathbf{F}_{\mathrm{diag}}[\hat{\mathbf{H}}']\mathbf{O}^T)^{-1}$. Hence, the sensitivity bound is attained by maximizing a diagonal Fisher matrix in the transformed modes $\hat{\mathbf{H}}'=\mathbf{O}\hat{\mathbf{H}}$ and then transforming to the original modes with the orthogonal matrix $\mathbf{O}$. The transformation $\mathbf{O}$ is in general not local in the modes and as such completely changes the correlation properties. Notice that the transformed modes need not necessarily be decomposable into a well-separated tensor product structure since they are obtained by mixing the original modes.

Despite the diagonal form of $\mathbf{F}_{\mathrm{diag}}[\hat{\mathbf{H}}']$, the final matrix $\mathbf{O}\mathbf{F}_{\mathrm{diag}}[\hat{\mathbf{H}}']\mathbf{O}^T$ is generally non-diagonal and is only reached by a quantum state that is strongly entangled in the modes $\hat{\mathbf{H}}$. For example, for the special case $\mathbf{W}=\mathbf{n}\mathbf{n}^T$ we have $w_k=\delta_{k1}$, which in~(\ref{eq:genericG}) reduces to the problem of a single mode $\hat{H}'_1=\mathbf{n}\cdot\hat{\mathbf{H}}$. For a total number of $N$ particles, the Fisher information can reach values up to $F_Q[\hat{\rho},\hat{H}'_1]\leq (\delta\Lambda'_1)^2$, where $\delta\Lambda'_1\equiv\Lambda'_{1+}-\Lambda'_{1-}$ and $\Lambda'_{1\pm}$ are the largest and smallest eigenvalues of $\hat{H}'_1$, respectively. Denoting the respective eigenvectors by $|\Lambda'_{1\pm}\rangle$, this bound is saturated by the optimal state $(|\Lambda'_{1+}\rangle+|\Lambda'_{1-}\rangle)/\sqrt{2}$. By distinguishing the cases of positive and negative components $n_k$, we find $\delta\Lambda'_1=\sum_{k=1}^M|n_k|(\delta \Lambda_k)$ and $|\Lambda'_{1\pm}\rangle=\bigotimes_{k=1}^M|\Lambda_{k\pm\epsilon_k}\rangle$ with $\epsilon_k=\mathrm{sgn}(n_k)$. As expected, we recover our results on the Heisenberg limit~(\ref{eq:HLbnd}) in this case since $\mathbf{n}^T\boldsymbol{\Sigma}\mathbf{n}\geq\mathbf{n}^T\boldsymbol{\Sigma}_{\max}^{\mathbf{n}\mathbf{n}^T}\mathbf{n}=(\delta\Lambda'_1)^{-2}=(\mathbf{n}^T\mathbf{f}^{\mathbf{n}})^{-2}=(\mathbf{n}^T\mathbf{F}_{\mathrm{HL,b}}^{\mathbf{n}}\mathbf{n})^{-1}$ with $\mathbf{F}_{\mathrm{HL,b}}^{\mathbf{n}}=\mathbf{f}^{\mathbf{n}}\mathbf{f}^{\mathbf{n}T}$ and $\mathbf{f}^{\mathbf{n}}=(\epsilon_1\delta\Lambda_1,\dots,\epsilon_M\delta \Lambda_M)$. When expressed in the original modes, the optimal state takes on the form~(\ref{eq:psin}).


\begin{thebibliography}{100}
\bibitem{nota4}
Here and in the following the phase of the superposition states does not influence the sensitivity and is chosen as $e^{i\varphi}=1$.
\end{thebibliography}
\end{document}